\documentclass[aps,a4paper,superscriptaddress,preprintnumbers,showpacs,amsmath,amssymb]{revtex4}
\usepackage{graphicx}
\usepackage{dcolumn}
\usepackage{color}
\usepackage{latexsym,amsfonts}
\usepackage{textcomp}
\usepackage{bm}

\usepackage{ulem}

\begin{document}

 \title{On the structure \\ of the correlation coefficients $S(E_e)$
   and $U(E_e)$ of the neutron beta decay}

 \author{A. N. Ivanov}\email{ivanov@kph.tuwien.ac.at}
 \affiliation{Atominstitut, Technische Universit\"at Wien,
   Stadionallee 2, A-1020 Wien, Austria}
 \author{R.~H\"ollwieser}\email{roman.hoellwieser@gmail.com}
 \affiliation{Atominstitut, Technische Universit\"at Wien,
   Stadionallee 2, A-1020 Wien, Austria}\affiliation{Department of
   Physics, New Mexico State University, Las Cruces, New Mexico 88003,
   USA} \author{N. I. Troitskaya}\email{natroitskaya@yandex.ru}
 \affiliation{Atominstitut, Technische Universit\"at Wien,
   Stadionallee 2, A-1020 Wien, Austria}
 \author{M. Wellenzohn}\email{max.wellenzohn@gmail.com}\affiliation{Atominstitut,
   Technische Universit\"at Wien, Stadionallee 2, A-1020 Wien,
   Austria}\affiliation{FH Campus Wien, University of Applied
   Sciences, Favoritenstra\ss e 226, 1100 Wien, Austria}
 \author{Ya. A. Berdnikov}\email{berdnikov@spbstu.ru}\affiliation{
   Peter the Great St. Petersburg Polytechnic University,
   Polytechnicheskaya 29, 195251, Russian Federation}

\date{\today}

\begin{abstract}
In the standard effective $V - A$ theory of low-energy weak
interactions (i.e. in the Standard Model (SM)) we analyze the
structure of the correlation coefficients $S(E_e)$ and $U(E_e)$, where
$E_e$ is the electron energy. These correlation coefficients were
introduced to the electron-energy and angular distribution of the
neutron beta decay by Ebel and Feldman ( Nucl. Phys. {\bf 4}, 213
(1957)) in addition to the set of correlation coefficients proposed by
Jackson {\it et al.}  (Phys. Rev. {\bf 106}, 517 (1957)). The
correlation coefficients $S(E_e)$ and $U(E_e)$ are induced by
simultaneous correlations of the neutron and electron spins and
electron and antineutrino 3-momenta. These correlation structures do
no violate discrete P, C and T symmetries. We analyze the
contributions of the radiative corrections of order $O(\alpha/\pi)$,
taken to leading order in the large nucleon mass $m_N$ expansion, and
corrections of order $O(E_e/m_N)$, caused by weak magnetism and proton
recoil. In addition to the obtained SM corrections we calculate the
contributions of interactions beyond the SM (BSM contributions) in
terms of the phenomenological coupling constants of BSM interactions
by Jackson {\it et al.}  (Phys. Rev. {\bf 106}, 517 (1957)) and the
{\it second class} currents by Weinberg (Phys. Rev.{\bf 112}, 1375
(1958)).
\end{abstract}
 \maketitle

\section{Introduction}
\label{sec:introduction}

The general form of the electron-energy and angular distribution of
the neutron beta decay for polarized neutrons, polarized electrons and
unpolarized protons were proposed by Jackson {\it et al.}
\cite{Jackson1957a} and Ebel and Feldman \cite{Ebel1957}. In the
notations of Ref. \cite{Ivanov2021} it looks like
\begin{eqnarray}\label{eq:1}
\hspace{-0.15in}&&\frac{d^5 \lambda_n(E_e, \vec{k}_e,
  \vec{k}_{\bar{\nu}}, \vec{\xi}_n, \vec{\xi}_e)}{dE_e d\Omega_e
  d\Omega_{\bar{\nu}}} \propto \zeta(E_e)\,\Big\{1 +
b(E_e)\,\frac{m_e}{E_e} + a(E_e)\,\frac{\vec{k}_e\cdot
  \vec{k}_{\bar{\nu}}}{E_e E_{\bar{\nu}}} +
A(E_e)\,\frac{\vec{\xi}_n\cdot \vec{k}_e}{E_e} + B(E_e)\,
\frac{\vec{\xi}_n\cdot \vec{k}_{\bar{\nu}}}{E_{\bar{\nu}}} \nonumber\\
\hspace{-0.15in}&& + K_n(E_e)\,\frac{(\vec{\xi}_n\cdot
  \vec{k}_e)(\vec{k}_e\cdot \vec{k}_{\bar{\nu}})}{E^2_e
  E_{\bar{\nu}}}+ Q_n(E_e)\,\frac{(\vec{\xi}_n\cdot
  \vec{k}_{\bar{\nu}})(\vec{k}_e\cdot \vec{k}_{\bar{\nu}})}{ E_e
  E^2_{\bar{\nu}}} + D(E_e)\,\frac{\vec{\xi}_n\cdot (\vec{k}_e\times
  \vec{k}_{\bar{\nu}})}{E_e E_{\bar{\nu}}} + G(E_e)\,\frac{\vec{\xi}_e
  \cdot \vec{k}_e}{E_e} \nonumber\\
\hspace{-0.15in}&& + H(E_e)\,\frac{\vec{\xi}_e \cdot
  \vec{k}_{\bar{\nu}}}{E_{\bar{\nu}}} + N(E_e)\,\vec{\xi}_n\cdot
\vec{\xi}_e + Q_e(E_e)\,\frac{(\vec{\xi}_n\cdot \vec{k}_e)(
  \vec{k}_e\cdot \vec{\xi}_e)}{(E_e + m_e) E_e} +
K_e(E_e)\,\frac{(\vec{\xi}_e\cdot \vec{k}_e)( \vec{k}_e\cdot
  \vec{k}_{\bar{\nu}})}{(E_e + m_e)E_e E_{\bar{\nu}}} \nonumber\\
\hspace{-0.15in}&& + R(E_e)\,\frac{\vec{\xi}_n\cdot(\vec{k}_e \times
  \vec{\xi}_e)}{E_e} + L(E_e)\,\frac{\vec{\xi}_e\cdot(\vec{k}_e \times
  \vec{k}_{\bar{\nu}})}{E_eE_{\bar{\nu}}} +
S(E_e)\,\frac{(\vec{\xi}_n\cdot \vec{\xi}_e)(\vec{k}_e \cdot
  \vec{k}_{\bar{\nu}})}{E_e E_{\bar{\nu}}} +
T(E_e)\,\frac{(\vec{\xi}_n \cdot \vec{k}_{\bar{\nu}})(\vec{\xi}_e
  \cdot \vec{k}_e)}{E_e E_{\bar{\nu}}} \nonumber\\
\hspace{-0.15in}&&+ U(E_e)\, \frac{(\vec{\xi}_n\cdot
  \vec{k}_e)(\vec{\xi}_e \cdot \vec{k}_{\bar{\nu}})}{E_e
  E_{\bar{\nu}}} + V(E_e)\, \frac{\vec{\xi}_n\cdot (\vec{\xi}_e \times
  \vec{k}_{\bar{\nu}})}{E_{\bar{\nu}}} + W(E_e)\,
\frac{\vec{\xi}_n\cdot (\vec{k}_e \times
  \vec{k}_{\bar{\nu}})(\vec{\xi}_e \cdot \vec{k}_e)}{(E_e + m_e) E_e
  E_{\bar{\nu}}} \Big\},
\end{eqnarray}
where $\vec{\xi}_n$ and $\vec{\xi}_e$ are unit 3-vectors of
spin-polarizations of the neutron and electron, $(E_e, \vec{k}_e)$ and
$(E_{\bar{\nu}}, \vec{k}_{\bar{\nu}})$ are energies and 3-momenta of
the electron and antineutrino, $d\Omega_e$ and $d\Omega_{\bar{\nu}}$
are infinitesimal solid angles in directions of 3-momenta of the
electron and antineutrino, respectively.

The analysis of the distribution in Eq.(\ref{eq:1}) within the
standard effective $V - A$ theory of low-energy weak interactions
\cite{Feynman1958, Marshak1969, Abele2008, Nico2009} (i.e within the
Standard Model (SM)), carried out to leading order in the large
nucleon mass $m_N$ expansion \cite{Ivanov2021}, has shown that
correlation coefficients $a(E_e)$, $A(E_e)$, $B(E_e)$, $G(E_e)$,
$H(E_e)$, $N(E_e)$, $Q_e(E_e)$ and $K_e(E_e)$ of the electron-energy
and angular distribution by Jackson {\it et al.}  \cite{Jackson1957a}
and the correlation coefficient $T(E_e)$, introduced by Ebel and
Feldman \cite{Ebel1957}, survive and depend on the axial coupling
constant $g_A$ only \cite{Abele2018, Sirlin2018, PDG2020}, which
appears in the effective $V - A$ theory of low-energy weak
interactions by renormalization of the hadronic axial-vector current
by strong low-energy interactions \cite{Marshak1969, DeAlfaro1973}.
The function $\zeta(E_e)$ defines the contributions of different
corrections to the neutron lifetime \cite{Ivanov2013}. In the SM it is
equal to unity in the leading order of the large nucleon mass $m_N$
expansion and at the neglect of radiative corrections \cite{Abele2008,
  Nico2009} (see also \cite{Ivanov2013}).  In Refs. \cite{Sirlin1967,
  Shann1971, Ando2004, Gudkov2006} (see also \cite{Ivanov2013}) the
radiative corrections of order $O(\alpha/\pi)$ (or so-called {\it
  outer} model-independent radiative corrections \cite{Wilkinson1970})
were calculated to leading order in the large nucleon mass $m_N$
expansion to the neutron lifetime and correlation coefficients
$a(E_e)$, caused by electron-antineutron 3-momentum correlations,
$A(E_e)$ and $B(E_e)$, defining the electron- and antineutrino
asymmetries, respectively.  In turn, the {\it outer} radiative
corrections of order $O(\alpha/\pi)$ were calculated to leading order
in the large nucleon mass $m_N$ expansion to the correlation
coefficients $G(E_e)$, $H(E_e)$, $N(E_e)$, $Q_e(E_e)$, and $K_e(E_e)$
in \cite{Ivanov2017, Ivanov2019a} and to the correlation coefficient
$T(E_e)$ in \cite{Ivanov2021}. These correlation coefficients are
induced by correlations of the electron spin with a neutron spin and
3-momenta of the electron and antineutrino. The corrections of order
$O(E_e/m_N)$, caused by weak magnetism and proton recoil, were
calculated i) to the neutron lifetime and correlation coefficients
$a(E_e)$, $A(E_e)$ and $B(E_e)$ in \cite{Bilenky1959, Wilkinson1982}
(see also \cite{Ando2004, Gudkov2006, Ivanov2013}), ii) to the
correlation coefficients $G(E_e)$, $H(E_e)$, $N(E_e)$, $Q_e(E_e)$ and
$K_e(E_e)$ in \cite{Ivanov2017, Ivanov2019a} and iii) to the
correlation coefficient $T(E_e)$ in \cite{Ivanov2021}. The correlation
coefficients $D(E_e)$, $R(E_e)$ and $L(E_e)$, characterizing the
strength of violation of time reversal invariance (T-odd effect)
\cite{Itzykson1980}, are induced by the distortion of the Dirac wave
function of the decay electron in the Coulomb field of the decay
proton \cite{Jackson1957b, Jackson1958, Callan1967, Ando2009} (see
also \cite{Ivanov2019a}). The correlation coefficient $b(E_e)$ is the
Fierz interference term \cite{Fierz1937}. It is assumed that the Fierz
interference term is caused by interactions beyond the SM
\cite{Fierz1937}. As regards the contemporary experimental and
theoretical status of the Fierz interference term we refer to
Refs. \cite{Hardy2020} - \cite{Ivanov2020}. So one may conclude that
the neutron lifetime and the correlation coefficients of the
electron-energy and angular distribution of the neutron beta decay
proposed by Jackson {\it et al.} \cite{Jackson1957a} are investigated
theoretically well in the SM at the level of $10^{-4} - 10^{-3}$,
caused by the {\it outer} radiative corrections of order
$O(\alpha/\pi)$ and the corrections of order $O(E_e/m_N)$, induced by
weak magnetism and proton recoil.

This paper is addressed to the analysis of the structure of the
correlation coefficients $S(E_e)$ and $U(E_e)$, introduced by Ebel and
Feldman \cite{Ebel1957}. As has been shown in \cite{Ivanov2021} these
correlation coefficients do not survive to leading order in the large
nucleon mass $m_N$ expansion in contrast to the correlation
coefficient $T(E_e)$.

The paper is organized as follows. In section \ref{sec:SM} we adduce
the analytical expressions for the correlation coefficients $S(E_e)$
and $U(E_e)$ in dependence of i) the radiative corrections of order
$O(\alpha/\pi)$, calculated to leading order in the large nucleon mass
$m_N$ expansion, and ii) the corrections of order $O(E_e/m_N)$, caused
by weak magnetism and proton recoil. In section \ref{sec:BSM} we give
the contributions of interactions beyond the SM, expressed in terms of
the phenomenological coupling constants of the effective
phenomenological BSM interactions by Jackson {\it et al.}
\cite{Jackson1957a} and the contributions of the {\it second class}
currents by Weinberg \cite{Weinberg1958}. In section
\ref{sec:Abschluss} we give the total expressions for the $S(E_e)$ and
$U(E_e)$. We discuss the obtained results and the usage of these
correlation coefficients for experimental searches of interactions
beyond the SM.  We point out that the obtained SM theoretical
background of the correlation coefficients $S(E_e)$ and $U(E_e)$ at
the level a few parts of $10^{-4}$ should be very useful for
experimental searches of contributions of interactions beyond the SM
in the experiments with transversally polarized decay electrons
\cite{Bodek2016}. In
Appendices A and B of the Supplemental Material within the SM we give
in details the calculations of the correlation coefficients $S(E_e)$
and $U(E_e)$ and the analysis of the correlation structure of the
neutron radiative beta decay for polarized neutrons, polarized
electrons, unpolarized protons and unpolarized photons.

\section{Correlation coefficients $S(E_e)$ and $U(E_e)$ in the
  Standard Model}
\label{sec:SM}

In the SM with the account for the contributions of the radiative
corrections of order $O(\alpha/\pi)$ and the corrections of order
$O(E_e/m_N)$, caused by weak magnetism and proton recoil, the neutron
beta decay can be described by the standard effective $V - A$
low-energy weak interaction \cite{Feynman1958, Marshak1969} and
electromagnetic interaction with the Lagrangian
\begin{eqnarray}\label{eq:2}
\hspace{-0.3in}{\cal L}_{\rm W\gamma}(x) = {\cal L}_{\rm W}(x) + {\cal
  L}_{\rm em}(x),
\end{eqnarray}
where ${\cal L}_{\rm W}(x)$ and ${\cal L}_{\rm em}(x)$ are the
Lagrangian of the standard effective $V - A$ low-energy weak
interactions \cite{Feynman1958, Marshak1969} (see also
\cite{Ivanov2013})
\begin{eqnarray}\label{eq:3}
\hspace{-0.3in}{\cal L}_{\rm W}(x) = -
G_V\,\Big\{[\bar{\psi}_p(x)\gamma_{\mu}(1 - g_A
  \gamma^5)\psi_n(x)] + \frac{\kappa}{2 m_N}
\partial^{\nu}[\bar{\psi}_p(x)\sigma_{\mu\nu}\psi_n(x)]\Big\}
        [\bar{\psi}_e(x)\gamma^{\mu}(1 - \gamma^5)\psi_{\nu}(x)]
\end{eqnarray}
and the Lagrangian of electromagnetic interactions \cite{Itzykson1980}
\begin{eqnarray}\label{eq:4}
\hspace{-0.3in}{\cal L}_{\rm em}(x) = -
e\big\{[\bar{\psi}_p(x)\gamma_{\mu}\psi_p(x)] -
[\bar{\psi}_e(x)\gamma^{\mu}\psi_e(x)]\big\} A_{\mu}(x),
\end{eqnarray}
respectively, where $G_V$ is the vector weak coupling constant,
including the Cabibbo-Kobayashi-Maskawa (CKM) matrix element $V_{ud}$
\cite{PDG2020}, $g_A$ is the real axial coupling constant
\cite{Abele2018, Sirlin2018}, $\psi_p(x)$, $\psi_n(x)$, $\psi_e(x)$
and $\psi_{\nu}(x)$ are the field operators of the proton, neutron,
electron and antineutrino, respectively, $\gamma^{\mu} = (\gamma^0,
\vec{\gamma}\,)$, $\gamma^5$ and $\sigma^{\mu\nu} =
\frac{i}{2}\,(\gamma^{\mu}\gamma^{\nu} - \gamma^{\nu} \gamma^{\mu})$
are the Dirac matrices \cite{Itzykson1980}; $\kappa = \kappa_p -
\kappa_n = 3.7059$ is the isovector anomalous magnetic moment of the
nucleon, defined by the anomalous magnetic moments of the proton
$\kappa_p = 1.7929$ and the neutron $\kappa_n = - 1.9130$ and measured
in nuclear magneton \cite{PDG2020}, and $m_N = (m_n + m_p)/2$ is the
average nucleon mass; $e$ is the electric charge of the proton, and
$A_{\mu}(x)$ is a 4-vector electromagnetic potential.

For the calculation of the correlation coefficients under
consideration we use the amplitude of the neutron beta decay,
calculated in \cite{Ivanov2013} (see also \cite{Ivanov2019a} and the
Supplemental Material). The detailed calculation we have
carried out in the Supplemental Material. Below we adduce only the
obtained results.

\subsection*{\bf Analytical expressions for the correlation
  coefficients $S(E_e)$ and $U(E_e)$ in the Standard Model}

In Eq.(\ref{eq:A.9}) of Appendix A in the Supplemental Material we
have defined the general expression for the structure part of the
electron-energy and angular distribution of the neutron beta decay for
a polarized neutron, a polarized electron and an unpolarized
proton. According to this expression, we have shown that the
contributions of the radiative corrections of order $O(\alpha/\pi)$,
caused by one-virtual photon exchanges \cite{Sirlin1967, Shann1971,
  Ando2004, Gudkov2006} (for the detailed calculations we refer to
\cite{Ivanov2013}) do not appear in the correlations coefficients
$S(E_e)$ and $U(E_e)$, respectively. In Appendix B of the Supplemental
Material we have shown that the neutron radiative beta decay $n \to p
+ e^- + \bar{\nu}_e + \gamma$ does not contribute to the correlation
coefficients $S(E_e)$ and $U(E_e)$. It is well-known \cite{Berman1958,
  Kinoshita1959, Berman1962, Kaellen1967, Abers1968} (see also
\cite{Sirlin1967, Shann1971} and \cite{Ivanov2013}) that the
contribution of the neutron radiative beta decay is extremely needed
for cancellation of the infrared divergences in the radiative
corrections of order $O(\alpha/\pi)$, caused by one-virtual photon
exchanges.

Thus (see Eq.(\ref{eq:A.9})) the contributions, caused by the SM
interactions, appear in the correlation coefficients $S(E_e)$ and
$U(E_e)$ only due to weak magnetism and proton recoil. For the
correlation coefficients $\zeta(E_e)^{(\rm SM)}S(E_e)^{(\rm SM)} $ and
$\zeta(E_e)^{(\rm SM)} U(E_e)^{(\rm SM)} $ we have obtained the
following analytical expressions
\begin{eqnarray}\label{eq:5}
 \hspace{-0.3in}\zeta(E_e)^{(\rm SM)} S(E_e)^{(\rm SM)}  &=& \frac{1}{1 + 3
  g^2_A}\,\frac{m_e}{m_N}\,\big(- 5 g^2_A - g_A (\kappa - 4) + (\kappa
+ 1)\big), \nonumber\\
\hspace{-0.3in}\zeta(E_e)^{(\rm SM)} U(E_e)^{(\rm SM)} &=& 0,
\end{eqnarray}
where for the calculation of the corrections of order $O(E_e/m_N)$,
caused weak magnetism and proton recoil, we have taken into account
the contribution of the phase-volume of the neutron beta decay (see
Eq.(\ref{eq:A.3})). The correlation function $\zeta(E_e)^{(\rm SM)} $
was calculated in \cite{Gudkov2006, Ivanov2013}. It is equal to unity
at the neglect of the contributions of radiative corrections and
corrections, caused by weak magnetism and proton recoil. Hence, the
correlation coefficients $S(E_e)^{(\rm SM)} $ and $U(E_e)^{(\rm SM)}$,
including the SM contributions of order $O(E_e/m_N)$, are equal to
\begin{eqnarray}\label{eq:6}
\hspace{-0.3in}S(E_e)^{(\rm SM)}  &=& \frac{1}{1 + 3
  g^2_A}\,\frac{m_e}{m_N}\,\big(- 5 g^2_A - g_A(\kappa - 4) + (\kappa
+ 1)\big), \nonumber\\
\hspace{-0.3in}U(E_e)^{(\rm SM)} &=& 0.
\end{eqnarray}
Now we may move on to calculating the contributions of interactions
beyond the SM.

\section{Contributions of interactions beyond the Standard Model and
  {\it second class} currents of the G-odd correlations}
\label{sec:BSM}

For the calculation of the contributions of interactions beyond the SM
we use the effective phenomenological Lagrangian of BSM interactions
proposed by Jackson {\it et al.}  \cite{Jackson1957a} (see also
\cite{Herczeg2001, Severijns2006}). In turn, the account for the
contributions of the {\it second class} currents or the $G$--odd
correlations (see \cite{Lee1956a}) we follow Weinberg
\cite{Weinberg1958}, Gardner and Zhang \cite{Gardner2001}, and Gardner
and Plaster \cite{Gardner2013} (see also \cite{Ivanov2018,
  Ivanov2019a, Ivanov2021}). Skipping intermediate calculations we
give the results
 \begin{eqnarray}\label{eq:7}
\hspace{-0.3in}S(E_e)^{(\rm BSM)} &=& \frac{1}{1 + 3 g^2_A}\,{\rm
  Re}(C_T - \bar{C}_T) - \big({\rm Re}g_2(0) - {\rm
  Re}f_3(0)\big)\,\,\frac{2 g_A}{1 + 3 g^2_A}\, \frac{m_e}{m_N},
\nonumber\\
\hspace{-0.3in}U(E_e)^{(\rm BSM)} &=& - \frac{1}{1 + 3 g^2_A}\,{\rm
  Re}(C_T - \bar{C}_T) + \big({\rm Re}g_2(0) - g_A{\rm
  Re}f_3(0)\,\frac{2}{1 + 3 g^2_A}\, \frac{m_e}{m_N},
 \end{eqnarray}
where $C_T$ and $\bar{C}_T$ are the phenomenological tensor coupling
constants of the effective phenomenological BSM interactions by
Jackson {\it et al.} \cite{Jackson1957a}, and ${\rm Re}f_3(0)$ and
${\rm Re}g_2(0)$ are the phenomenological coupling constants of the
induced scalar and tensor {\it second class} currents
\cite{Weinberg1958, Gardner2001, Gardner2013} (see also
\cite{Hardy2020}), respectively.  The contributions of the tensor BSM
interactions by Jackson {\it et al.}  \cite{Jackson1957a} are linear
in the phenomenological tensor coupling constants $C_T$ and
$\bar{C}_T$. This agrees well with the result obtained by Ebel and
Feldman \cite{Ebel1957}. However, in addition to the result obtained
by Ebel and Feldman \cite{Ebel1957} we, following
\cite{Bhattacharya2012, Gardner2012, Cirigliano2012, Cirigliano2013a,
  Cirigliano2013b} (see also \cite{Ivanov2013, Ivanov2017, Ivanov2018,
  Ivanov2019a, Ivanov2021}), have taken the contributions of the
phenomenological vector coupling constants $C_V$ and $\bar{C}_V$ in
the linear approximation, i.e. $C_V = 1 + \delta C_V$ and $\bar{C}_V =
- 1 + \delta \bar{C}_V$, where we have used the notations of
\cite{Ivanov2013, Ivanov2017, Ivanov2018, Ivanov2019a, Ivanov2021}.

\section{Discussion}
\label{sec:Abschluss}

We have analyzed the structure of the correlation coefficients
$S(E_e)$ and $U(E_e)$, introduced by Ebel and Feldman \cite{Ebel1957}
in addition to the set of correlation coefficients proposed by Jackson
{\it et al.} \cite{Jackson1957a}.  Summing up the SM contributions,
caused by weak magnetism and proton recoil only, and contributions
beyond the SM we obtain the following expressions
\begin{eqnarray}\label{eq:8}
\hspace{-0.3in}S(E_e) &=& \frac{1}{1 + 3
  g^2_A}\,\frac{m_e}{m_N}\,\big(- 5 g^2_A - g_A (\kappa - 4) + (\kappa
+ 1)\big) + \frac{1}{1 + 3 g^2_A}\,{\rm Re}(C_T -
\bar{C}_T)\nonumber\\ \hspace{-0.3in}&-& \big({\rm Re}g_2(0) - {\rm
  Re}f_3(0)\big)\,\frac{2 g_A}{1 + 3 g^2_A}\, \frac{m_e}{m_N},\nonumber\\
\hspace{-0.3in}U(E_e) &=& - \frac{1}{1 + 3 g^2_A}\,{\rm Re}(C_T -
\bar{C}_T) + \big({\rm Re}g_2(0) - g_A {\rm Re}f_3(0)\big)\,\frac{2}{1
  + 3 g^2_A}\, \frac{m_e}{m_N}.
\end{eqnarray}
For the axial couping constant $g_A = 1.2764$
\cite{Abele2018} the correlation coefficients $S(E_e)$ and $U(E_e)$
are given by
\begin{eqnarray}\label{eq:9}
\hspace{-0.3in}S(E_e) &=& - 2.83 \times 10^{-4} + 0.17\,\Big({\rm Re}(C_T
- \bar{C}_T) + 1.39 \times 10^{-3}\, {\rm Re}f_3(0)\Big) - 2.36 \times
10^{-4}\, {\rm Re}g_2(0) , \nonumber\\
\hspace{-0.3in}U(E_e) &=& - 0.17 \,\Big({\rm Re}(C_T - \bar{C}_T) +
1.39\times 10^{-3}\, {\rm Re}f_3(0)\Big) + 1.85 \times 10^{-4}\, {\rm
  Re}g_2(0),
\end{eqnarray}
where we have also used $m_e = 0.5110\,{\rm MeV}$ and $m_N = (m_n +
m_p)/2 = 938.9188\, {\rm MeV}$ \cite{PDG2020}. 

We would like to notice that the correlation structures of the
correlation coefficients $S(E_e)$ and $U(E_e)$ and as well as the
correlation coefficients $T(E_e)$ are even with respect to parity
transformation (P-even), charge conjugation (C-even) and time reversal
transformation (T-even). However, in contrast to the correlation
coefficient $T(E_e)$, the absolute value of which is of about
$|T(E_e)| \sim 1$, the absolute values of the correlation coefficients
$S(E_e)$ and $U(E_e)$ are of a few orders of magnitude smaller. It is
also important to mention that unlike the correlation coefficient
$T(E_e)$ the correlation coefficients $S(E_e)$ and $U(E_e)$ do not
depend on the electron energy $E_e$.

The correlation coefficients $S(E_e)$ and $U(E_e)$ can, in principle,
be investigated in experiments with both longitudinally and
transversally polarized decay electrons \cite{Bodek2016} (see also
\cite{Ivanov2021}). However, a successful results for searches of
interactions beyond the SM one may expect only from experiments with
experimental uncertainties of about a few parts of $10^{-5}$. In this
case any deviation of the correlation coefficient $S(E_e)$ from $-
2.83\times 10^{-4}$, caused by weak magnetism and proton recoil,
should testify a presence of interactions beyond the SM. Since most
likely $f_3(0) = 0$ \cite{Hardy2020, Severijns2019} (see also
\cite{Hardy2009}) the contributions of the phenomenological tensor BSM
interactions by Jackson {\it et al.}  \cite{Jackson1957a},
proportional to ${\rm Re}(C_T - \bar{C}_T)$, can be distinguished from
the contributions of the tensor {\it second class} current, defined by
the phenomenological tensor coupling constant ${\rm Re}g_2(0)$, only
after the measurement of the contribution of the phenomenological
tensor {\it second class} currents to the correlation coefficient
$T(E_e)$ \cite{Ivanov2021}. In case of $f_3(0) \neq 0$ and if in the
neutron beta decay the absolute value of the Fierz interference term
$b$ could be of order $10^{-2}$ (see, for example, \cite{Abele2019,
  Ivanov2019y}) after the measurement of the phenomenological tensor
coupling constant ${\rm Re}g_2(0)$ from the correlation coefficient
$T(E_e)$, the contribution of the scalar coupling constant ${\rm
  Re}f_3(0)$ to the correlation coefficients $S(E_e)$ and $U(E_e)$
could be screened by the contributions of the phenomenological tensor
coupling constants ${\rm Re}(C_T - \bar{C}_T)$ of the phenomenological
tensor BSM interactions by Jackson {\it et al.}  \cite{Jackson1957a}.

\section{Acknowledgements}

We thank Hartmut Abele for discussions stimulating this work.  The
work of A. N. Ivanov was supported by the Austrian ``Fonds zur
F\"orderung der Wissenschaftlichen Forschung'' (FWF) under contracts
P31702-N27 and P26636-N20, and ``Deutsche F\"orderungsgemeinschaft''
(DFG) AB 128/5-2. The work of R. H\"ollwieser was supported by the
Deutsche Forschungsgemeinschaft in the SFB/TR 55. The work of
M. Wellenzohn was supported by the MA 23 (FH-Call 16) under the
project ``Photonik - Stiftungsprofessur f\"ur Lehre''.

\newpage

\section{\bf the Supplemental Material}
\label{sec:appendix}

\section*{Appendix A: The electron-energy and angular
  distribution of the neutron beta decay for polarized neutrons,
  polarized electrons and unpolarized protons}
\renewcommand{\theequation}{A-\arabic{equation}}
\setcounter{equation}{0}

Following \cite{Ivanov2013, Ivanov2017, Ivanov2019a} (see also
\cite{Ivanov2021}) we define the electron-energy and angular
distribution of the neutron beta decay for a polarized neutron, a
polarized electron, and an unpolarized proton as follows
\begin{eqnarray}\label{eq:A.1}
\hspace{-0.3in}\frac{d^5\lambda_{\beta^-_c\gamma}(E_e,\vec{k}_e,
  \vec{k}_{\bar{\nu}},\vec{\xi}_n, \vec{\xi}_e)}{d E_e
  d\Omega_ed\Omega_{\bar{\nu}}} &=& (1 + 3 g^2_A) \,\frac{|G_V|^2}{16
  \pi^5}\,(E_0 - E_e )^2\,\sqrt{E^2_e - m^2_e}\,E_e\,F(E_e, Z =
1)\nonumber\\
\hspace{-0.3in}&&\times \, \Phi_n(\vec{k}_e, \vec{k}_{\bar{\nu}})
\sum_{\rm pol.}\frac{|M(n \to p e^- \bar{\nu}_e)|^2}{(1 + 3
  g^2_A)|G_V|^2 64 m^2_n E_e E_{\bar{\nu}}},
\end{eqnarray}
where the sum is over polarizations of massive fermions. Then, $F(E_e,
Z = 1)$ is the relativistic Fermi function, describing the
electron--proton final--state Coulomb interaction, is equal to (see,
for example, \cite{Blatt1952} (see also \cite{ Wilkinson1982} and a
discussion in \cite{Ivanov2017})
\begin{eqnarray}\label{eq:A.2}
\hspace{-0.3in}F(E_e, Z = 1 ) = \Big(1 +
\frac{1}{2}\gamma\Big)\,\frac{4(2 r_pm_e\beta)^{2\gamma}}{\Gamma^2(3 +
  2\gamma)}\,\frac{\displaystyle e^{\,\pi \alpha/\beta}}{(1 -
  \beta^2)^{\gamma}}\,\Big|\Gamma\Big(1 + \gamma + i\,\frac{\alpha
}{\beta}\Big)\Big|^2,
\end{eqnarray}
where $\beta = k_e/E_e = \sqrt{E^2_e - m^2_e}/E_e$ is the electron
velocity, $\gamma = \sqrt{1 - \alpha^2} - 1$, $r_p$ is the electric
radius of the proton \cite{Antognini2013}. The
function $\Phi_n(\vec{k}_e, \vec{k}_{\bar{\nu}})$ defines the
contribution of the phase-volume of the neutron beta decay
\cite{Ivanov2013, Ivanov2020b}. It is equal to \cite{Ivanov2013,
  Ivanov2020b}
\begin{eqnarray}\label{eq:A.3}
\hspace{-0.3in}\Phi_n(\vec{k}_e, \vec{k}_{\bar{\nu}}) = 1 + 3
\frac{E_e}{m_N}\Big(1 - \frac{\vec{k}_e\cdot \vec{k}_{\bar{\nu}}}{E_e
  E_{\bar{\nu}}}\Big),
\end{eqnarray}
taken to next-to-leading order in the large nucleon mass $m_N$
expansion. The amplitude of the neutron beta decay $M(n \to p e^-
\bar{\nu}_e)$, taking into account the contribution of the
corrections, caused by one-virtual photon exchanges, weak magnetism
and proton recoil, was calculated in \cite{Ivanov2013} (see also
\cite{Ivanov2019a}). It is given by
\begin{eqnarray}\label{eq:A.4}
\hspace{-0.3in}&&M(n \to p\,e^- \,\bar{\nu}_e) = - 2m_n\,G_V\Big\{\Big(1 +
       \frac{\alpha}{2\pi}\,f_{\beta^-_c}(E_e,\mu)\Big)[\varphi^{\dagger}_p
         \varphi_n][\bar{u}_e\,\gamma^0(1 -
         \gamma^5)v_{\bar{\nu}}]\nonumber\\
\hspace{-0.3in}&& + \tilde{g}_A \Big(1 +
\frac{\alpha}{2\pi}\,f_{\beta^-_c}(E_e,\mu)\Big)[\varphi^{\dagger}_p
  \vec{\sigma}\,\varphi_n]\cdot [\bar{u}_e \vec{\gamma}\,(1 -
  \gamma^5)v_{\bar{\nu}}] -
\frac{\alpha}{2\pi}\,g_F(E_e)\,[\varphi^{\dagger}_p
  \varphi_n][\bar{u}_e\,(1 -
  \gamma^5)v_{\bar{\nu}}]\nonumber\\\hspace{-0.3in}&& -
\frac{\alpha}{2\pi}\,\tilde{g}_A g_F(E_e) [\varphi^{\dagger}_p
  \vec{\sigma}\,\varphi_n]\cdot [\bar{u}_e \gamma^0\vec{\gamma}\,(1 -
  \gamma^5)v_{\bar{\nu}}] - \frac{m_e}{2
  m_N}\,[\varphi^{\dagger}_p\varphi_n][\bar{u}_e\,(1 -
  \gamma^5)v_{\bar{\nu}}]\nonumber\\
\hspace{-0.3in}&& - \frac{\tilde{g}_A}{2
  m_N}[\varphi^{\dagger}_p(\vec{\sigma}\cdot \vec{k}_p) \varphi_n
]\,[\bar{u}_e\,\gamma^0 (1 - \gamma^5)v_{\bar{\nu}}] - i\,
\frac{\kappa + 1}{2 m_N} [\varphi^{\dagger}_p (\vec{\sigma}\times
  \vec{k}_p) \varphi_n] \cdot [\bar{u}_e\,\vec{\gamma}\,(1 -
  \gamma^5)v_{\bar{\nu}}] \Big\},
\end{eqnarray}
where $\varphi_p$ and $\varphi_n$ are Pauli spinorial wave functions
of the proton and neutron, $u_e$ and $v_{\nu}$ are Dirac wave
functions of the electron and electron antineutrino, $\vec{\sigma}$
are the Pauli $2\times 2$ matrices, and $\tilde{g}_A = g_A (1 - E_0/2
m_N)$, $E_0 = (m^2_n - m^2_p + m^2_e)/2 m_n = 1.2926\, {\rm MeV}$ is
the end-point energy of the electron-energy spectrum of the neutron
beta decay \cite{Abele2008, Nico2009, PDG2020}, and $\vec{k}_p = -
\vec{k}_e - \vec{k}_{\nu}$ is the proton 3--momentum in the rest frame
of the neutron. The functions $f_{\beta^-_c}(E_e,\mu)$ and $g_F(E_e)$
were calculated by Sirlin \cite{Sirlin1967} (see also Eq.(D-51) of
Ref. \cite{Ivanov2013} and ), $\mu$ is a covariant infrared cut-off
introduced as a finite virtual photon mass \cite{Sirlin1967} (see also
\cite{Berman1958, Kinoshita1959, Berman1962, Kaellen1967,
  Abers1968}). The function $g_F(E_e)$ (see Eq.(D-44) of
Ref. \cite{Ivanov2013}) is equal to
\begin{eqnarray}\label{eq:A.5}
\hspace{-0.3in} g_F(E_e) = \frac{\sqrt{1 - \beta^2}}{2 \beta}\, {\ell n}\Big(\frac{1 + \beta}{1 - \beta}\Big).
\end{eqnarray}
It is defined by the contributions of one-virtual-photon exchanges
\cite{Sirlin1967} (see also \cite{Ivanov2013}).  Using
Eq.(\ref{eq:A.4}) for the square of the absolute value of the
amplitude $M(n \to p e^- \bar{\nu}_e)$, summed over polarizations of
massive fermions, we obtain the following expression
\begin{eqnarray*}%\label{eq:A.6}
\hspace{-0.3in}&&\sum_{\rm pol.}\frac{|M(n \to p e^-
  \bar{\nu}_e)|^2}{(1 + 3 g^2_A)|G_V|^2 64 m^2_n E_e E_{\bar{\nu}}} =
\frac{1}{(1 + 3 g^2_A) 8 E_e E_{\bar{\nu}}}\Big\{\Big(1 +
\frac{\alpha}{\pi}\,f_{\beta^-_c}(E_e,\mu)\Big) \Big({\rm tr}\{(1 +
  \vec{\xi}_n\cdot \vec{\sigma}\,)\}\,{\rm tr}\{(\hat{k}_e + m_e
  \gamma^5 \hat{\zeta}_e) \gamma^0 \hat{k}_{\bar{\nu}}\gamma^0
  \nonumber\\
\hspace{-0.3in}&& \times (1 - \gamma^5)\} + \tilde{g}_A {\rm tr}\{(1 +
\vec{\xi}_n\cdot \vec{\sigma}\,) \vec{\sigma}\,\}\cdot {\rm
  tr}\{(\hat{k}_e + m_e \gamma^5 \hat{\zeta}_e) \big(\gamma^0
\hat{k}_{\bar{\nu}}\vec{\gamma} + \vec{\gamma}\, \hat{k}_{\bar{\nu}}
\gamma^0\big) (1 - \gamma^5)\} + \tilde{g}^2 _A {\rm tr}\{(1 +
\vec{\xi}_n\cdot \vec{\sigma}\,) \sigma^a\sigma^b \} \nonumber\\  
\hspace{-0.3in}&& \times {\rm tr}\{(\hat{k}_e + m_e \gamma^5
\hat{\zeta}_e) \gamma^b \hat{k}_{\bar{\nu}}\gamma^a (1 -
\gamma^5)\}\Big) - \Big(\frac{\alpha}{2\pi} g_F(E_e) + \frac{m_e}{2
  m_N}\Big) {\rm tr}\{(1 + \vec{\xi}_n\cdot \vec{\sigma}\,)\}\Big({\rm
  tr}\{(m_e + \hat{k}_e \gamma^5
\hat{\zeta}_e)\hat{k}_{\bar{\nu}}\gamma^0 (1 - \gamma^5)\} \nonumber\\
\end{eqnarray*}
\begin{eqnarray}\label{eq:A.6}
\hspace{-0.3in}&& + {\rm tr}\{(m_e + \hat{k}_e \gamma^5
\hat{\zeta}_e)\gamma^0 \hat{k}_{\bar{\nu}} (1 + \gamma^5)\}\Big) -
\tilde{g}_A \Big(\frac{\alpha}{2\pi}\,g_F(E_e) + \frac{m_e}{2
  m_N}\Big) {\rm tr}\{(1 + \vec{\xi}_n\cdot
\vec{\sigma}\,)\vec{\sigma}\,\} \cdot \Big({\rm tr}\{( m_e + \hat{k}_e
\gamma^5 \hat{\zeta}_e) \hat{k}_{\bar{\nu}} \vec{\gamma} (1 -
\gamma^5)\}  \nonumber\\
\hspace{-0.3in}&& + {\rm tr}\{(\hat{k}_e + m_e \gamma^5 \hat{\zeta}_e)
\vec{\gamma}\, \hat{k}_{\bar{\nu}} (1 + \gamma^5)\}\Big) - \tilde{g}_A
\frac{\alpha}{2\pi} g_F(E_e) {\rm tr}\{(1 + \vec{\xi}_n\cdot
\vec{\sigma}\,) \vec{\sigma}\,\}\cdot \Big({\rm tr}\{(m_e + \hat{k}_e
\gamma^5 \hat{\zeta}_e) \gamma^0 \vec{\gamma}\, \hat{k}_{\bar{\nu}}
\gamma^0 (1 - \gamma^5)\} \nonumber\\
\hspace{-0.3in}&& - {\rm tr}\{(m_e + \hat{k}_e \gamma^5 \hat{\zeta}_e)
\gamma^0 \hat{k}_{\bar{\nu}} \gamma^0 \vec{\gamma}\, (1 +
\gamma^5)\}\Big) - \tilde{g}^2_A \frac{\alpha}{2\pi}\,g_F(E_e)\, {\rm
  tr}\{(1 + \vec{\xi}_n\cdot \vec{\sigma}\,) \sigma^a \sigma^b\}
\Big({\rm tr}\{(m_e + \hat{k}_e \gamma^5 \hat{\zeta}_e) \gamma^0
\gamma^b \hat{k}_{\bar{\nu}} \gamma^a (1 - \gamma^5)\} \nonumber\\
\hspace{-0.3in}&& - {\rm tr}\{(m_e + \hat{k}_e \gamma^5 \hat{\zeta}_e)
\gamma^b \hat{k}_{\bar{\nu}} \gamma^0 \gamma^a (1 + \gamma^5)\} -
\frac{\tilde{g}_A}{m_N}{\rm tr}\{(1 + \vec{\xi}_n\cdot
\vec{\sigma}\,)(\vec{\sigma} \cdot \vec{k}_p)\}\,{\rm tr}\{(\hat{k}_e
+ m_e \gamma^5 \hat{\zeta}_e) \gamma^0 \hat{k}_{\bar{\nu}}\gamma^0 (1
- \gamma^5)\} - \frac{\tilde{g}^2_A}{2 m_N}\nonumber\\
\hspace{-0.3in}&& \times \Big( {\rm tr}\{(1 + \vec{\xi}_n\cdot
\vec{\sigma}\,)\vec{\sigma}\,(\vec{\sigma}\cdot \vec{k}_p)\} \cdot
    {\rm tr}\{(\hat{k}_e + m_e \gamma^5 \hat{\zeta}_e) \gamma^0
    \hat{k}_{\bar{\nu}}\vec{\gamma}\, (1 - \gamma^5)\} + {\rm tr}\{(1
    + \vec{\xi}_n\cdot \vec{\sigma}\,)(\vec{\sigma}\cdot
    \vec{k}_p)\vec{\sigma}\,\} \cdot {\rm tr}\{(\hat{k}_e + m_e
    \gamma^5 \hat{\zeta}_e) \vec{\gamma}\,
    \hat{k}_{\bar{\nu}}\gamma^0 \nonumber\\
\hspace{-0.3in}&& \times (1 - \gamma^5)\}\Big) - i \,\frac{\kappa +
  1}{2 m_N}{\rm tr}\{(1 +
\vec{\xi}_n\cdot\vec{\sigma})(\vec{\sigma}\times \vec{k}_p)\} \cdot
    {\rm tr}\{(\hat{k}_e + m_e \gamma^5 \hat{\zeta}_e)
    \big(\vec{\gamma}\,\hat{k}_{\bar{\nu}} \gamma^0 -
    \gamma^0\hat{k}_{\bar{\nu}} \vec{\gamma}\big)(1 - \gamma^5)\} \nonumber\\
\hspace{-0.3in}&& - i\,\tilde{g}_A \frac{\kappa + 1}{2 m_N}\Big({\rm
  tr}\{(1 + \vec{\xi}_n\cdot\vec{\sigma})
\big(\sigma^a(\vec{\sigma}\times \vec{k}_p)^b - (\vec{\sigma}\times
\vec{k}_p)^a\sigma^b\big)\}\,{\rm tr}\{(\hat{k}_e + m_e \gamma^5
\hat{\zeta}_e) \gamma^b\,\hat{k}_{\bar{\nu}}\gamma^a (1 -
\gamma^5)\}\Big\},
\end{eqnarray}
where $\zeta_e$ is the 4--vector of the spin--polarization of the
electron. It is defined by \cite{Itzykson1980}
\begin{eqnarray}\label{eq:A.7}
\zeta_e = (\zeta^0_e, \vec{\zeta}_e) = \Big(\frac{\vec{k}_e\cdot
  \vec{\xi}_e}{m_e}, \vec{\xi}_e + \frac{(\vec{k}_e \cdot \vec{\xi}_e)
  \vec{k}_e }{m_e(E_e + m_e)}\Big).
\end{eqnarray}
The 4-vector $\zeta_e$ of the spin--polarization of the electron is
normalized by $\zeta^2_e = - 1$ and obeys also the constraint
$k_e\cdot \zeta_e = 0$ \cite{Itzykson1980}. Calculating the traces
over the nucleon degrees of freedom and using the properties of the
Dirac matrices \cite{Itzykson1980}
\begin{eqnarray}\label{eq:A.8}
\hspace{-0.3in}\gamma^{\alpha}\gamma^{\nu}\gamma^{\mu} =
\gamma^{\alpha}\eta^{\nu\mu} - \gamma^{\nu}\eta^{\mu\alpha} +
\gamma^{\mu}\eta^{\alpha\nu} +
i\,\varepsilon^{\alpha\nu\mu\beta}\,\gamma_{\beta}\gamma^5,
\end{eqnarray}
where $\eta^{\mu\nu}$ is the metric tensor of the Minkowski
space--time, $\varepsilon^{\alpha\nu\mu\beta}$ is the Levi--Civita
tensor defined by $\varepsilon^{0123} = 1$ and
$\varepsilon_{\alpha\nu\mu\beta}= - \varepsilon^{\alpha\nu\mu\beta}$
\cite{Itzykson1980}, we transcribe the right-hand-side (r.h.s.) of
Eq.(\ref{eq:A.6}) into the form \cite{Ivanov2013, Ivanov2017,
  Ivanov2019a} (see also \cite{Ivanov2021})
\begin{eqnarray}\label{eq:A.9}
\hspace{-0.3in}&&\sum_{\rm pol.}\frac{|M(n \to p e^- \bar{\nu}_e)|^2}{(1
  + 3 g^2_A)|G_V|^2 64 m^2_n E_e E_{\bar{\nu}}} = \frac{1 + 3
  \tilde{g}^2_A}{(1 + 3 g^2_A) 4 E_e}\Big\{\Big(1 +
\frac{\alpha}{\pi}\,f_{\beta^-_c}(E_e,\mu)\Big) \Big[\Big(1 +
  \tilde{B}_0\frac{\vec{\xi}_n \cdot
    \vec{k}_{\bar{\nu}}}{E_{\bar{\nu}}}\Big){\rm tr}\{(\hat{k}_e + m_e
  \gamma^5 \hat{\zeta}_e) \gamma^0 (1 - \gamma^5)\} \nonumber\\
\hspace{-0.3in}&&+ \Big(\tilde{A}_0 \vec{\xi}_n + \tilde{a}_0
\frac{\vec{k}_{\bar{\nu}}}{E_{\bar{\nu}}}\Big) \cdot {\rm
  tr}\{(\hat{k}_e + m_e \gamma^5 \hat{\zeta}_e) \vec{\gamma}\, (1 -
\gamma^5)\} \Big] - \frac{1}{1 + 3 \tilde{g}^2_A}
\Big(\frac{\alpha}{\pi} g_F(E_e) + \frac{m_e}{ m_N}\Big) \Big({\rm
  tr}\{(m_e + \hat{k}_e \gamma^5 \hat{\zeta}_e)\} \nonumber\\
\hspace{-0.3in}&& - \frac{\vec{k}_{\bar{\nu}}}{E_{\bar{\nu}}}\cdot {\rm
  tr}\{(m_e + \hat{k}_e \gamma^5 \hat{\zeta}_e) \gamma^0
\vec{\gamma}\, \gamma^5\}\Big) - \frac{\tilde{g}_A}{1 + 3
  \tilde{g}^2_A} \Big(\frac{\alpha}{\pi}\,g_F(E_e) +
\frac{m_e}{m_N}\Big) \Big(\frac{\vec{\xi}_n \cdot
  \vec{k}_{\bar{\nu}}}{E_{\bar{\nu}}} {\rm tr}\{( m_e + \hat{k}_e
\gamma^5 \hat{\zeta}_e)\} \nonumber\\
\hspace{-0.3in}&& - \vec{\xi}_n \cdot {\rm tr}\{(m_e + \hat{k}_e
\gamma^5 \hat{\zeta}_e)\gamma^0 \vec{\gamma}\,\gamma^5\} + i\,
\frac{\vec{\xi}_n \times \vec{k}_{\bar{\nu}}}{E_{\bar{\nu}}} \cdot
     {\rm tr}\{(m_e + \hat{k}_e \gamma^5 \hat{\zeta}_e)\gamma^0
     \vec{\gamma}\,\}\Big) - \frac{\tilde{g}_A}{1 + 3 \tilde{g}^2_A}
     \frac{\alpha}{\pi} g_F(E_e) \nonumber\\
\hspace{-0.3in}&& \times \Big( \frac{\vec{\xi}_n\cdot
  \vec{k}_{\bar{\nu}}}{E_{\bar{\nu}}}\,{\rm tr}\{(m_e + \hat{k}_e
\gamma^5 \hat{\zeta}_e)\} - \vec{\xi}_n \cdot {\rm tr}\{(m_e +
\hat{k}_e \gamma^5 \hat{\zeta}_e) \gamma^0 \vec{\gamma}\,\gamma^5\} -
i\,\frac{\vec{\xi}_n\times \vec{k}_{\bar{\nu}}}{E_{\bar{\nu}}} \cdot
{\rm tr}\{(m_e + \hat{k}_e \gamma^5 \hat{\zeta}_e) \gamma^0
\vec{\gamma}\,\}  \nonumber\\
\hspace{-0.3in}&& - \frac{\tilde{g}^2_A}{1 + 3 \tilde{g}^2_A}
\frac{\alpha}{\pi}\,g_F(E_e)\Big[\Big(3 + 2\, \frac{\vec{\xi}_n\cdot
    \vec{k}_{\bar{\nu}}}{E_{\bar{\nu}}}\Big){\rm tr}\{(m_e + \hat{k}_e
  \gamma^5 \hat{\zeta}_e)\} + \Big(2 \vec{\xi}_n +
  \frac{\vec{k}_{\bar{\nu}}}{E_{\bar{\nu}}}\Big)\cdot {\rm tr}\{(m_e +
  \hat{k}_e \gamma^5 \hat{\zeta}_e) \gamma^0 \vec{\gamma}\,
  \gamma^5\}\Big] \nonumber\\
\hspace{-0.3in}&&- \frac{\tilde{g}_A}{1 + 3 \tilde{g}^2_A}\,
\frac{1}{m_N}\,\Big((\vec{\xi}_n \cdot
  \vec{k}_p)\,{\rm tr}\{(\hat{k}_e + m_e \gamma^5
\hat{\zeta}_e) \gamma^0 (1 - \gamma^5)\} + (\vec{\xi}_n \cdot
\vec{k}_p)\frac{\vec{k}_{\bar{\nu}}}{E_{\bar{\nu}}}\cdot {\rm
  tr}\{(\hat{k}_e + m_e \gamma^5 \hat{\zeta}_e) \vec{\gamma}\, (1 -
\gamma^5)\}\Big) - \frac{\tilde{g}^2_A}{1 + 3 \tilde{g}^2_A}\,
\frac{1}{m_N} \nonumber\\
\hspace{-0.3in}&& \times \, \Big[\frac{\vec{k}_p \cdot
    \vec{k}_{\bar{\nu}}}{E_{\bar{\nu}}}\, {\rm tr}\{(\hat{k}_e + m_e
  \gamma^5 \hat{\zeta}_e) \gamma^0 (1 - \gamma^5)\} + \Big(\vec{k}_p +
  \frac{\vec{\xi}_n \cdot \vec{k}_{\bar{\nu}}}{E_{\bar{\nu}}}\vec{k}_p
  - \frac{\vec{k}_p \cdot
    \vec{k}_{\bar{\nu}}}{E_{\bar{\nu}}}\,\vec{\xi}_n\Big) \cdot {\rm
    tr}\{(\hat{k}_e + m_e \gamma^5 \hat{\zeta}_e) \vec{\gamma}\, (1 -
  \gamma^5)\}\Big] - \frac{\kappa + 1}{1 + 3 \tilde{g}^2_A}\,
\frac{1}{m_N} \nonumber\\
\hspace{-0.3in}&& \times \, \Big(\frac{\vec{\xi}_n \cdot
  \vec{k}_{\bar{\nu}}}{E_{\bar{\nu}}}\vec{k}_p - \frac{\vec{k}_p \cdot
  \vec{k}_{\bar{\nu}}}{E_{\bar{\nu}}}\,\vec{\xi}_n\Big) \cdot {\rm
  tr}\{(\hat{k}_e + m_e \gamma^5 \hat{\zeta}_e) \vec{\gamma}\, (1 -
\gamma^5)\} + \frac{\tilde{g}_A}{1 + 3 \tilde{g}^2_A}\, \frac{\kappa +
  1}{m_N}\Big[\Big(2\,\frac{\vec{k}_p \cdot
    \vec{k}_{\bar{\nu}}}{E_{\bar{\nu}}} + 2 (\vec{\xi}_n \cdot
  \vec{k}_p)\Big) \,{\rm tr}\{(\hat{k}_e + m_e \gamma^5 \hat{\zeta}_e) \nonumber\\
\hspace{-0.3in}&& \times \, \gamma^0 (1 - \gamma^5)\} + \Big(- 2
\vec{k}_p - \frac{\vec{\xi}_n \cdot
  \vec{k}_{\bar{\nu}}}{E_{\bar{\nu}}}\vec{k}_p - \frac{\vec{k}_p \cdot
  \vec{k}_{\bar{\nu}}}{E_{\bar{\nu}}}\,\vec{\xi}_n\Big) \cdot {\rm
  tr}\{(\hat{k}_e + m_e \gamma^5 \hat{\zeta}_e) \vec{\gamma}\, (1 -
\gamma^5)\}\Big]\Big\},
\end{eqnarray}
where $\tilde{a}_0$, $\tilde{A}_0$ and $\tilde{B}_0$ are defined in
terms of the axial coupling constant $\tilde{g}_A$ 
\begin{eqnarray}\label{eq:A.10}
\hspace{-0.3in} a_0 = \frac{1 - \tilde{g}^2_A}{1 + 3
  \tilde{g}^2_A}\quad,\quad A_0 = 2 \frac{\tilde{g}_A(1 -
  \tilde{g}_A)}{1 + 3 \tilde{g}^2_A}\quad,\quad B_0 = 2
\frac{\tilde{g}_A(1 + \tilde{g}_A)}{1 + 3 \tilde{g}^2_A}.
\end{eqnarray}
Before the calculation of the traces over leptonic degrees of freedom
one may see that the terms proportional two $g_F(E_e)
i(\vec{\xi}_n\times \vec{k}_{\bar{\nu}})\cdot {\rm tr}\{(m_e +
\hat{k}_e \gamma^5 \hat{\zeta}_e) \gamma^0 \vec{\gamma}\,\}$, which
are responsible for contributions of the radiative corrections of
order $O(\alpha/\pi)$ to the correlation coefficients $S(E_e)$ and
$U(E_e)$, cancel each other out. Hence, there are no contributions of
the radiative corrections of order $O(\alpha/\pi)$, caused by
one-virtual-photon exchanges, to the correlation coefficients $S(E_e)$
and $U(E_e)$, respectively.

In Eq.(\ref{eq:A.9}) the second term on the third line from above,
proportional to $m_e/m_N$, and last four lines define the
contributions of order $O(E_e/m_N)$ of weak magnetism and proton
recoil to the correlation coefficients of the neutron beta decay.
Having calculated the traces over leptonic degrees of freedom, taking
into account the contribution of the phase-volume Eq.(\ref{eq:A.3})
and keeping only the contributions with the correlation structures,
inducing the correlation coefficients $S(E_e)$ and $U(E_e)$, we obtain
the SM corrections, caused by weak magnetism and proton recoil only,
which we give in Eq.(\ref{eq:5}).

\section*{Appendix B: The electron-photon-energy and angular
  distribution of the neutron radiative beta decay for polarized
  neutrons, polarized electrons and unpolarized protons and photons}
\renewcommand{\theequation}{B-\arabic{equation}}
\setcounter{equation}{0}

Following \cite{Ivanov2013, Ivanov2017, Ivanov2019a, Ivanov2021} we
define the electron-photon-energy and angular distribution of the
neutron radiative beta decay for a polarized neutron, a polarized
electron, a polarized photon and an unpolarized proton as follows
\begin{eqnarray}\label{eq:B.1}
\hspace{-0.3in}\frac{d^8\lambda_{\beta^-_c\gamma}(E_e,\omega,\vec{k}_e,
  \vec{k}_{\bar{\nu}},\vec{q},\vec{\xi}_n, \vec{\xi}_e)_{\lambda'
    \lambda}}{d\omega d E_e
  d\Omega_ed\Omega_{\bar{\nu}}\Omega_{\gamma}} &=& (1 + 3 g^2_A) \,
\frac{\alpha}{\pi}\,\frac{|G_V|^2}{(2\pi)^6}\,\sqrt{E^2_e -
  m^2_e}\,E_e\,F(E_e, Z = 1)\,(E_0 - E_e -
\omega)^2\,\frac{1}{\omega}\nonumber\\
\hspace{-0.3in}&&\times\,\sum_{\rm pol.}\frac{|M(n \to p e^-
  \bar{\nu}_e \gamma)|^2_{\lambda' \lambda} \omega^2}{(1 + 3 g^2_A) e^2|G_V|^2
  64 m^2_n E_e E_{\bar{\nu}}},
\end{eqnarray}
where we sum over polarizations of massive fermions. Since we
calculate the contribution of the neutron radiative beta decay to
leading order in the large nucleon mass $m_N$ expansion, the
contribution of the phase volume of the decay is equal to unity.  The
photon state is determined by the 4--momentum $q^{\mu} = (\omega,
\vec{q}\,)$ and the 4-vector of polarization
$\varepsilon^{\mu}(q)_{\lambda}$ with $\lambda = 1,2$, obeying the
constraints $\varepsilon^*(q)_{\lambda'}\cdot \varepsilon_{\lambda}(q)
= - \delta_{\lambda'\lambda}$ and $q \cdot \varepsilon_{\lambda}(q) =
0$. In the tree-approximation and to leading order in the large
nucleon mass $m_N$ expansion the amplitude of the neutron radiative
beta decay is equal to \cite{Ivanov2013}
\begin{eqnarray}\label{eq:B.2}
M(n\to p\,e^- \bar{\nu}_e\gamma)_{\lambda} = e\,
G_V\frac{m_n}{\omega}\,\frac{1}{E_e - \vec{n}\cdot \vec{k}_e}\Big\{
[\varphi^{\dagger}_p \varphi_n] [\bar{u}_e Q_{\lambda} \gamma^0 (1 -
  \gamma^5)v_{\bar{\nu}}] + g_A
[\varphi^{\dagger}_p\vec{\sigma}\,\varphi_n]\cdot [\bar{u}_e Q_{\lambda}
  \vec{\gamma}\,(1 - \gamma^5)v_{\bar{\nu}}]\Big\}.
\end{eqnarray}
The hermitian conjugate amplitude is determined by
\begin{eqnarray}\label{eq:B.3}
\hspace{-0.3in}M^{\dagger}(n\to p\,e^- \bar{\nu}_e\gamma)_{\lambda'} = e\,
G_V\frac{m_n}{\omega}\,\frac{1}{E_e - \vec{n}\cdot \vec{k}_e}\Big\{
[\varphi^{\dagger}_p \varphi_n] [\bar{u}_e \bar{Q}_{\lambda'} \gamma^0 (1 -
  \gamma^5)v_{\bar{\nu}}] + g_A
[\varphi^{\dagger}_p\vec{\sigma}\,\varphi_n]\cdot [\bar{u}_e \bar{Q}_{\lambda'}
  \vec{\gamma}\,(1 - \gamma^5)v_{\bar{\nu}}]\Big\},
\end{eqnarray}
where $\vec{n} = \vec{q}/\omega$, $Q = 2(\varepsilon^*\cdot k_e) +
\hat{\varepsilon}^* \hat{q}$ and $\bar{Q} = \gamma^0 Q^{\dagger}
\gamma^0 = 2(\varepsilon\cdot k_e) + \hat{q}\hat{\varepsilon}$. Then,
$\varphi_n$ and $\varphi_p$ are the Pauli wave functions of the
neutron and proton, $u_e$ and $v_{\bar{\nu}}$ are the Dirac wave
functions of the electron and antineutrino, respectively. The sum over
polarizations of the massive fermions is equal to \cite{Ivanov2013,
  Ivanov2017, Ivanov2019a}
\begin{eqnarray}\label{eq:B.4}
\hspace{-0.15in}&&\sum_{\rm pol.}\frac{|M(n \to p e^- \bar{\nu}_e
  \gamma)|^2_{\lambda' \lambda} \omega^2}{(1 + 3 g^2_A) e^2|G_V|^2 64
  m^2_n E_e E_{\bar{\nu}}} = \frac{1}{(E_e - \vec{n}\cdot
  \vec{k}_e)^2}\, \frac{1}{(1 + 3 g^2_A) 32 E_e
  E_{\bar{\nu}}}\Big\{{\rm tr}\{(1 + \vec{\xi}_n\cdot
\vec{\sigma}\,)\}\,{\rm tr}\{(\hat{k}_e + m_e \gamma^5 \hat{\zeta}_e)
Q_{\lambda} \gamma^0 \hat{k}_{\bar{\nu}}\gamma^0
\bar{Q}_{\lambda'}\nonumber\\
\hspace{-0.15in}&&\times\, (1 - \gamma^5)\} + g_A {\rm tr}\{(1 +
\vec{\xi}_n\cdot \vec{\sigma}\,) \vec{\sigma}\,\}\cdot {\rm
  tr}\{(\hat{k}_e + m_e \gamma^5 \hat{\zeta}_e) Q_{\lambda} \gamma^0
\hat{k}_{\bar{\nu}}\vec{\gamma}\, \bar{Q}_{\lambda'} (1 - \gamma^5)\}
+ g_A {\rm tr}\{(1 + \vec{\xi}_n\cdot \vec{\sigma}\,)
\vec{\sigma}\,\}\cdot {\rm tr}\{(\hat{k}_e + m_e \gamma^5
\hat{\zeta}_e) \nonumber\\
\hspace{-0.15in}&&\times \, Q_{\lambda} \vec{\gamma}\,
\hat{k}_{\bar{\nu}}\gamma^0 \bar{Q}_{\lambda'} (1 - \gamma^5)\} +
g^2_A {\rm tr}\{(1 + \vec{\xi}_n\cdot \vec{\sigma}\,) \sigma^a\sigma^b
\} {\rm tr}\{(\hat{k}_e + m_e \gamma^5 \hat{\zeta}_e)
Q_{\lambda}\gamma^b \hat{k}_{\bar{\nu}}\gamma^a\, \bar{Q}_{\lambda'}
(1 - \gamma^5)\}\Big\}.
\end{eqnarray}
Having calculated the traces over the nucleon degrees of freedom and
using the properties of the Dirac matrices Eq.(\ref{eq:A.8}) we
transcribe the r.h.s. of Eq.(\ref{eq:B.4}) into the form
\cite{Ivanov2013, Ivanov2017, Ivanov2019a}
\begin{eqnarray}\label{eq:B.5}
\hspace{-0.15in}&&\sum_{\rm pol.}\frac{|M(n \to p e^- \bar{\nu}_e
  \gamma)|^2_{\lambda' \lambda} \omega^2}{(1 + 3 g^2_A) e^2|G_V|^2 64
  m^2_n E_e E_{\bar{\nu}}} = \frac{1}{(E_e - \vec{n}\cdot
  \vec{k}_e)^2}\, \frac{1}{ 16 E_e}\Big\{\Big(1 + B_0\,
\frac{\vec{\xi}_n\cdot \vec{k}_{\bar{\nu}}}{E_{\bar{\nu}}}\Big){\rm
  tr}\{(\hat{k}_e + m_e \gamma^5 \hat{\zeta}_e) Q_{\lambda} \gamma^0
\bar{Q}_{\lambda'} (1 - \gamma^5)\} \nonumber\\
\hspace{-0.15in}&& + \Big(A_0\, \vec{\xi}_n +
a_0\,\frac{\vec{k}_{\bar{\nu}}}{E_{\bar{\nu}}}\Big) \cdot {\rm
  tr}\{(\hat{k}_e + m_e \gamma^5 \hat{\zeta}_e) Q_{\lambda}
\vec{\gamma}\, \bar{Q}_{\lambda'} (1 - \gamma^5)\}\Big\}.
\end{eqnarray}
The traces over Dirac matrices in Eq.(\ref{eq:B.5}) were calculated in
the covariant form in \cite{Ivanov2013, Ivanov2019a}. The result is
\begin{eqnarray}\label{eq:B.6}
\hspace{-0.3in}&&\frac{1}{16}\,{\rm tr}\{\hat{a}\, Q_{\lambda}
\gamma^{\mu} \bar{Q}_{\lambda'}(1 - \gamma^5)\} =
(\varepsilon^*_{\lambda}\cdot k_e)(\varepsilon_{\lambda'}\cdot k_e)
a^{\mu} + \frac{1}{2}\,\Big( (\varepsilon^*_{\lambda}\cdot
k_e)(\varepsilon_{\lambda'}\cdot a) + (\varepsilon^*_{\lambda}\cdot
a)(\varepsilon_{\lambda'}\cdot k_e) - (\varepsilon^*_{\lambda}\cdot
\varepsilon_{\lambda'}) (a\cdot q) \Big) q^{\mu} \nonumber\\
\hspace{-0.3in}&& - \frac{1}{2}\,\Big((\varepsilon^*_{\lambda}\cdot
k_e) \varepsilon^{\mu}_{\lambda'} + \varepsilon^{*\mu}_{\lambda}
(\varepsilon_{\lambda'}\cdot k_e)\Big) (a \cdot q) -
i\,\frac{1}{2}\,\varepsilon^{\mu\nu\alpha\beta}\Big((\varepsilon^*_{\lambda}\cdot
k_e) \varepsilon_{\lambda' \nu} - \varepsilon^*_{\lambda
  \nu}(\varepsilon_{\lambda'}\cdot k_e)\Big) a_{\alpha} q_{\beta} -
i\,\frac{1}{2}\,q^{\mu} \varepsilon^{\rho\varphi\alpha\beta}
\varepsilon^*_{\lambda \rho} \varepsilon_{\lambda' \varphi}
a_{\alpha}q_{\beta},\nonumber\\
\hspace{-0.3in}&&
\end{eqnarray}
where $a = k_e$ or $a = \zeta_e$, respectively. As a result, for the
r.h.s. of Eq.(\ref{eq:B.5}) we obtain the following expression
\begin{eqnarray}\label{eq:B.7}
\hspace{-0.21in}&&\sum_{\rm pol.}\frac{|M(n \to p e^- \bar{\nu}_e
  \gamma)|^2_{\lambda' \lambda} \omega^2}{(1 + 3 g^2_A) e^2|G_V|^2 64
  m^2_n E_e E_{\bar{\nu}}} = \frac{1}{(E_e - \vec{n}\cdot
  \vec{k}_e)^2}\,\bigg\{\Big(1 + B_0\, \frac{\vec{\xi}_n\cdot
  \vec{k}_{\bar{\nu}}}{E_{\bar{\nu}}}\Big)\Big\{
\Big[(\varepsilon^*_{\lambda}\cdot k_e) (\varepsilon_{\lambda'}\cdot
k_e)\Big(1 + \frac{\omega}{E_e}\Big) -
\frac{1}{2}\,(\varepsilon^*_{\lambda}\cdot
\varepsilon_{\lambda'})\nonumber\\
\hspace{-0.3in}&&\times\, (k_e\cdot q) \,\frac{\omega}{E_e} -
\frac{1}{2}\,\Big((\varepsilon^*_{\lambda}\cdot k_e)
\varepsilon^{0}_{\lambda'} + \varepsilon^{* 0}_{\lambda}
(\varepsilon_{\lambda'}\cdot k_e)\Big) \frac{k_e \cdot q}{E_e}\Big] -
\frac{m_e}{E_e}\Big[(\varepsilon^*_{\lambda}\cdot
  k_e)(\varepsilon_{\lambda'}\cdot k_e) \zeta^0_e + \frac{1}{2}\,\Big(
  (\varepsilon^*_{\lambda}\cdot k_e)(\varepsilon_{\lambda'}\cdot
  \zeta_e) + (\varepsilon^*_{\lambda}\cdot \zeta_e) \nonumber\\
\hspace{-0.3in}&&\times\, (\varepsilon_{\lambda'}\cdot k_e) -
(\varepsilon^*_{\lambda}\cdot \varepsilon^*_{\lambda'}) (\zeta_e\cdot
q) \Big) \omega - \frac{1}{2}\,\Big((\varepsilon^*_{\lambda}\cdot k_e)
\varepsilon^{0}_{\lambda'} + \varepsilon^{* 0}_{\lambda}
(\varepsilon_{\lambda'}\cdot k_e)\Big) (\zeta_e \cdot q) \Big] \Big\}
+ \Big(A_0\, \vec{\xi}_n +
a_0\,\frac{\vec{k}_{\bar{\nu}}}{E_{\bar{\nu}}}\Big) \cdot
\Big\{\Big[(\varepsilon^*_{\lambda}\cdot k_e)(\varepsilon_{\lambda'}\cdot
k_e) \nonumber\\
\hspace{-0.3in}&&\times\,\frac{\vec{k}_e}{E_e} + \Big(
(\varepsilon^*_{\lambda}\cdot k_e)(\varepsilon_{\lambda'}\cdot k_e) -
\frac{1}{2}\,(\varepsilon^*_{\lambda}\cdot \varepsilon_{\lambda'})
(k_e\cdot q) \Big)\,\vec{n}\, \frac{\omega}{E_e} -
\frac{1}{2}\,\Big((\varepsilon^*_{\lambda}\cdot k_e)
\vec{\varepsilon}_{\lambda'} + \vec{\varepsilon}^{\; *}_{\lambda}
(\varepsilon_{\lambda'}\cdot k_e)\Big) \frac{k_e \cdot q}{E_e}\Big] -
\frac{m_e}{E_e}\Big[(\varepsilon^*_{\lambda}\cdot
  k_e)(\varepsilon_{\lambda'}\cdot k_e) \nonumber\\
\hspace{-0.3in}&&\times\, \vec{\zeta}_e + \frac{1}{2}\,\Big(
(\varepsilon^*_{\lambda}\cdot k_e)(\varepsilon_{\lambda'}\cdot
\zeta_e) + (\varepsilon^*_{\lambda}\cdot
\zeta_e)(\varepsilon_{\lambda'}\cdot k_e) -
(\varepsilon^*_{\lambda}\cdot \varepsilon_{\lambda'}) (\zeta_e\cdot q)
\Big)\, \omega\, \vec{n} - \frac{1}{2}\,
\Big((\varepsilon^*_{\lambda}\cdot k_e) \vec{\varepsilon}_{\lambda'} +
\vec{\varepsilon}^{\; *}_{\lambda} (\varepsilon_{\lambda'}\cdot
k_e)\Big) (\zeta_e \cdot q) \Big] \Big\}\bigg\}.\nonumber\\
\hspace{-0.3in}&&
\end{eqnarray}
The r.h.s. of Eq.(\ref{eq:B.7}) we calculate in the physical gauge
$\varepsilon_{\lambda} = (0, \vec{\varepsilon}_{\lambda})$
\cite{Ivanov2013, Ivanov2017, Ivanov2019a, Ivanov2017b}, where the
polarization vector $\vec{\varepsilon}_{\lambda}$ obeys the
constraints 
\begin{eqnarray}\label{eq:B.8}
\hspace{-0.3in}\vec{q}\cdot \vec{\varepsilon}^{\;*}_{\lambda} &=&
\vec{q}\cdot \vec{\varepsilon}_{\lambda'} =
0\;,\;\vec{\varepsilon}^{\;*}_{\lambda}\cdot
\vec{\varepsilon}_{\lambda'} = \delta_{\lambda \lambda'} \;,\;
\sum_{\lambda = 1,2}\vec{\varepsilon}^{\,i
  *}_{\lambda}\vec{\varepsilon}^{\,j}_{\lambda} = \delta^{ij} -
\frac{\vec{k}^{\,i} \vec{k}^{\,j}}{\omega^2} = \delta^{ij} -
\vec{n}^{\,i}\, \vec{n}^{\,j}\;,\;\sum_{j =1,2,3}\sum_{\lambda =
  1,2}\vec{\varepsilon}^{\,j
  *}_{\lambda}\vec{\varepsilon}^{\,j}_{\lambda} = 2.
\end{eqnarray}
In the physical gauge $\varepsilon_{\lambda} = (0,
\vec{\varepsilon}_{\lambda})$ we obtain for the r.h.s. of
Eq.(\ref{eq:B.7}) the following expression
\begin{eqnarray}\label{eq:B.9}
\hspace{-0.3in}&&\sum_{\rm pol.}\frac{|M(n \to p e^- \bar{\nu}_e
  \gamma)|^2_{\lambda' \lambda} \omega^2}{(1 + 3 g^2_A) e^2|G_V|^2 16
  m^2_n E_e E_{\bar{\nu}}} = \frac{1}{(E_e - \vec{n}\cdot
  \vec{k}_e)^2}\bigg\{\Big(1 + B_0\, \frac{\vec{\xi}_n\cdot
  \vec{k}_{\bar{\nu}}}{E_{\bar{\nu}}}\Big)\Big\{
\Big[(\vec{\varepsilon}^{\; *}_{\lambda}\cdot \vec{k}_e)
  (\vec{\varepsilon}_{\lambda'}\cdot \vec{k}_e)\Big(1 +
  \frac{\omega}{E_e}\Big) +
  \frac{1}{2}\,(\vec{\varepsilon}^{\;*}_{\lambda}\cdot
  \vec{\varepsilon}_{\lambda'})\nonumber\\
\hspace{-0.3in}&&\times\, (k_e\cdot q) \,\frac{\omega}{E_e}\Big] -
\frac{m_e}{E_e}\Big[(\vec{\varepsilon}^{\; *}_{\lambda}\cdot
  \vec{k}_e)(\vec{\varepsilon}_{\lambda'}\cdot \vec{k}_e) \zeta^0_e +
  \frac{1}{2}\,\Big( (\vec{\varepsilon}^{\; *}_{\lambda}\cdot
  \vec{k}_e)(\vec{\varepsilon}_{\lambda'}\cdot \vec{\zeta}_e) +
  (\vec{\varepsilon}^{\; *}_{\lambda}\cdot \vec{\zeta}_e)\,
  (\vec{\varepsilon}_{\lambda'}\cdot \vec{k}_e) +
  (\vec{\varepsilon}^{\; *}_{\lambda}\cdot
  \vec{\varepsilon}_{\lambda'}) (\zeta_e\cdot q) \Big) \omega \Big]
\Big\} \nonumber\\
\hspace{-0.3in}&& + \Big(A_0\, \vec{\xi}_n +
a_0\,\frac{\vec{k}_{\bar{\nu}}}{E_{\bar{\nu}}}\Big) \cdot
\Big\{\Big[(\vec{\varepsilon}^{\; *}_{\lambda}\cdot
  \vec{k}_e)(\vec{\varepsilon}_{\lambda'}\cdot \vec{k}_e)
  \,\frac{\vec{k}_e}{E_e} + \Big( (\vec{\varepsilon}^{\;
    *}_{\lambda}\cdot \vec{k}_e)(\vec{\varepsilon}_{\lambda'}\cdot
  \vec{k}_e) + \frac{1}{2}\,(\vec{\varepsilon}^{\; *}_{\lambda}\cdot
  \vec{\varepsilon}_{\lambda'}) (k_e\cdot q) \Big)\,\vec{n}\,
  \frac{\omega}{E_e} + \frac{1}{2}\,\Big((\vec{\varepsilon}^{\;
    *}_{\lambda}\cdot \vec{k}_e) \vec{\varepsilon}_{\lambda'} \nonumber\\
\hspace{-0.3in}&& + \vec{\varepsilon}^{\; *}_{\lambda}
(\vec{\varepsilon}_{\lambda'}\cdot \vec{k}_e)\Big) \frac{k_e \cdot
  q}{E_e}\Big] - \frac{m_e}{E_e}\Big[(\vec{\varepsilon}^{\;
    *}_{\lambda}\cdot \vec{k}_e)(\vec{\varepsilon}_{\lambda'} \cdot
  \vec{k}_e) \,\vec{\zeta}_e + \frac{1}{2}\,\Big(
  (\vec{\varepsilon}^{\; *}_{\lambda}\cdot
  \vec{k}_e)(\vec{\varepsilon}_{\lambda'}\cdot \vec{\zeta}_e) +
  (\vec{\varepsilon}^{\; *}_{\lambda}\cdot
  \vec{\zeta}_e)(\vec{\varepsilon}_{\lambda'}\cdot \vec{k}_e) +
  (\vec{\varepsilon}^{\; *}_{\lambda}\cdot
  \vec{\varepsilon}_{\lambda'}) (\zeta_e\cdot q) \Big) \, \omega\,
  \vec{n}\nonumber\\
\hspace{-0.3in}&& + \frac{1}{2}\, \Big((\vec{\varepsilon}^{\;
  *}_{\lambda}\cdot \vec{k}_e) \vec{\varepsilon}_{\lambda'} +
\vec{\varepsilon}^{\; *}_{\lambda} (\vec{\varepsilon}_{\lambda'}\cdot
\vec{k}_e)\Big) (\zeta_e \cdot q) \Big] \Big\}\bigg\}.
\end{eqnarray}
Plugging Eq.(\ref{eq:B.9}) into Eq.(\ref{eq:B.1}) we obtain the
electron-energy and angular distribution for a polarized neutron, a
polarized electron, an unpolarized proton and a polarized photon:
\begin{eqnarray}\label{eq:B.10}
\hspace{-0.15in}&&\frac{d^8\lambda_{\beta^-_c\gamma}(E_e,\omega,\vec{k}_e,
  \vec{k}_{\bar{\nu}},\vec{q},\vec{\xi}_n, \vec{\xi}_e)_{\lambda'
    \lambda}}{d\omega d E_e
  d\Omega_ed\Omega_{\bar{\nu}}\Omega_{\gamma}} = (1 + 3 g^2_A) \,
\frac{\alpha}{\pi}\,\frac{|G_V|^2}{(2\pi)^6}\,\sqrt{E^2_e -
  m^2_e}\,E_e\,F(E_e, Z = 1)\,(E_0 - E_e -
\omega)^2\,\frac{1}{\omega}\frac{1}{(E_e - \vec{n}\cdot \vec{k}_e)^2}
\nonumber\\
\hspace{-0.3in}&& \times \,\bigg\{\Big(1 + B_0\,
\frac{\vec{\xi}_n\cdot \vec{k}_{\bar{\nu}}}{E_{\bar{\nu}}}\Big)\Big\{
\Big[(\vec{\varepsilon}^{\; *}_{\lambda}\cdot \vec{k}_e)
  (\vec{\varepsilon}_{\lambda'}\cdot \vec{k}_e)\Big(1 +
  \frac{\omega}{E_e}\Big) +
  \frac{1}{2}\,(\vec{\varepsilon}^{\;*}_{\lambda}\cdot
  \vec{\varepsilon}_{\lambda'}) \, (k_e\cdot q)
  \,\frac{\omega}{E_e}\Big] -
\frac{m_e}{E_e}\Big[(\vec{\varepsilon}^{\; *}_{\lambda}\cdot
  \vec{k}_e)\, (\vec{\varepsilon}_{\lambda'}\cdot \vec{k}_e) \zeta^0_e
  + \frac{1}{2}\,\Big( (\vec{\varepsilon}^{\; *}_{\lambda}\cdot
  \vec{k}_e)\nonumber\\
\hspace{-0.3in}&& \times (\vec{\varepsilon}_{\lambda'}\cdot
\vec{\zeta}_e) + (\vec{\varepsilon}^{\; *}_{\lambda}\cdot
\vec{\zeta}_e) (\vec{\varepsilon}_{\lambda'}\cdot \vec{k}_e) +
(\vec{\varepsilon}^{\; *}_{\lambda}\cdot \vec{\varepsilon}_{\lambda'})
(\zeta_e\cdot q) \Big) \omega \Big] \Big\} + \Big(A_0 \vec{\xi}_n +
a_0 \frac{\vec{k}_{\bar{\nu}}}{E_{\bar{\nu}}}\Big) \cdot
\Big\{\Big[(\vec{\varepsilon}^{\; *}_{\lambda}\cdot
  \vec{k}_e)(\vec{\varepsilon}_{\lambda'}\cdot \vec{k}_e)
  \frac{\vec{k}_e}{E_e} + \Big( (\vec{\varepsilon}^{\;
    *}_{\lambda}\cdot \vec{k}_e)(\vec{\varepsilon}_{\lambda'}\cdot
  \vec{k}_e) \nonumber\\
\hspace{-0.3in}&& + \frac{1}{2}\,(\vec{\varepsilon}^{\;
  *}_{\lambda}\cdot \vec{\varepsilon}_{\lambda'}) (k_e\cdot q)
\Big)\,\vec{n}\, \frac{\omega}{E_e} +
\frac{1}{2}\,\Big((\vec{\varepsilon}^{\; *}_{\lambda}\cdot \vec{k}_e)
\vec{\varepsilon}_{\lambda'} + \vec{\varepsilon}^{\; *}_{\lambda}
(\vec{\varepsilon}_{\lambda'}\cdot \vec{k}_e)\Big) \frac{k_e \cdot
  q}{E_e}\Big] - \frac{m_e}{E_e}\Big[(\vec{\varepsilon}^{\;
    *}_{\lambda}\cdot \vec{k}_e)(\vec{\varepsilon}_{\lambda'} \cdot
  \vec{k}_e) \,\vec{\zeta}_e + \frac{1}{2}\,\Big(
  (\vec{\varepsilon}^{\; *}_{\lambda}\cdot
  \vec{k}_e)(\vec{\varepsilon}_{\lambda'}\cdot \vec{\zeta}_e)
  \nonumber\\
\hspace{-0.3in}&& + (\vec{\varepsilon}^{\; *}_{\lambda}\cdot
\vec{\zeta}_e)(\vec{\varepsilon}_{\lambda'}\cdot \vec{k}_e) +
(\vec{\varepsilon}^{\; *}_{\lambda}\cdot \vec{\varepsilon}_{\lambda'})
(\zeta_e\cdot q) \Big) \, \omega\, \vec{n} + \frac{1}{2}\,
\Big((\vec{\varepsilon}^{\; *}_{\lambda}\cdot \vec{k}_e)
\vec{\varepsilon}_{\lambda'} + \vec{\varepsilon}^{\; *}_{\lambda}
(\vec{\varepsilon}_{\lambda'}\cdot \vec{k}_e)\Big) (\zeta_e \cdot q)
\Big] \Big\}\bigg\}.
\end{eqnarray}
Summing up over polarizations of the photon we get
\begin{eqnarray}\label{eq:B.11}
\hspace{-0.15in}&&\frac{d^8\lambda_{\beta^-_c\gamma}(E_e,\omega,\vec{k}_e,
  \vec{k}_{\bar{\nu}},\vec{q},\vec{\xi}_n, \vec{\xi}_e)}{d\omega d E_e
  d\Omega_ed\Omega_{\bar{\nu}}\Omega_{\gamma}} = (1 + 3 g^2_A) \,
\frac{\alpha}{\pi}\,\frac{|G_V|^2}{(2\pi)^6}\,\sqrt{E^2_e -
  m^2_e}\,E_e\,F(E_e, Z = 1)\,(E_0 - E_e -
\omega)^2\,\frac{1}{\omega}
\nonumber\\
\hspace{-0.3in}&& \times \,\bigg\{\Big(1 + B_0\,
\frac{\vec{\xi}_n\cdot \vec{k}_{\bar{\nu}}}{E_{\bar{\nu}}}\Big)\Big\{
\Big[\frac{\beta^2 - (\vec{n}\cdot \vec{\beta}\,)^2}{(1 - \vec{n}\cdot
    \vec{\beta})^2} \Big(1 + \frac{\omega}{E_e}\Big) +
  \frac{\omega^2}{E^2_e}\,\frac{1}{1 - \vec{n}\cdot \vec{\beta}}\Big]
- \frac{m_e}{E_e}\Big[\frac{\beta^2 - (\vec{n} \cdot
    \vec{\beta}\,)^2}{(1 - \vec{n}\cdot \vec{\beta}\,)^2} \zeta^0_e +
  \frac{\omega}{E_e}\frac{\zeta^0_e -
    (\vec{n}\cdot \vec{\beta}\,)(\vec{n}\cdot \vec{\zeta}_e)}{(1 -
    \vec{n}\cdot \vec{\beta}\,)^2} \nonumber\\ 
\hspace{-0.3in}&& + \frac{\omega^2}{E^2_e} \frac{\zeta^0_e -
  \vec{n}\cdot \vec{\zeta}_e}{(1 - \vec{n}\cdot \vec{\beta}\,)^2}
\Big) \Big] \Big\} + \Big(A_0 \vec{\xi}_n + a_0
\frac{\vec{k}_{\bar{\nu}}}{E_{\bar{\nu}}}\Big) \cdot
\Big\{\Big[\frac{\beta^2 - (\vec{n} \cdot \vec{\beta}\,)^2}{(1 -
    \vec{n}\cdot \vec{\beta}\,)^2}\Big( \frac{\vec{k}_e}{E_e} +
  \vec{n}\, \frac{\omega}{E_e}\Big) + \frac{\omega}{E_e}\,
  \frac{\vec{\beta} - \vec{n}\,(\vec{n}\cdot \vec{\beta}\,)}{1 -
    \vec{n}\cdot \vec{\beta}} + \frac{\omega^2}{E^2_e}\,
  \frac{\vec{n}}{1 - \vec{n}\cdot \vec{\beta}}\Big]\nonumber\\
\hspace{-0.3in}&& - \frac{m_e}{E_e}\Big[\frac{\beta^2 - (\vec{n} \cdot
    \vec{\beta}\,)^2}{(1 - \vec{n}\cdot
    \vec{\beta}\,)^2}\,\vec{\zeta}_e +
  \frac{\omega}{E_e}\,\frac{\zeta^0_e - \vec{n}\cdot \vec{\zeta}_e}{(1
    - \vec{n}\cdot \vec{\beta}\,)^2}\,\vec{\beta} +
  \frac{\omega}{E_e}\,\frac{\vec{n}}{1 - \vec{n}\cdot
    \vec{\beta}}\,\zeta^0_e +
  \frac{\omega^2}{E^2_e}\,\frac{\vec{n}\,\big(\zeta^0_e - \vec{n}\cdot
    \vec{\zeta}_e\big)}{(1 - \vec{n}\cdot
    \vec{\beta}\,)^2}\Big\}\bigg\},
\end{eqnarray}
where we have used that $\vec{\beta}\cdot \vec{\zeta}_e = \zeta^0_e$.
The next step is to average over directions of the 3--momentum
$\vec{q} = \omega \vec{n}$ of the real photon. This gives
\begin{eqnarray}\label{eq:B.12}
\hspace{-0.15in}&&\frac{d^6\lambda_{\beta^-_c\gamma}(E_e,\omega,\vec{k}_e,
  \vec{k}_{\bar{\nu}}, \vec{\xi}_n, \vec{\xi}_e)}{d\omega d E_e
  d\Omega_ed\Omega_{\bar{\nu}}} = (1 + 3 g^2_A) \,
\frac{\alpha}{\pi}\,\frac{|G_V|^2}{16 \pi^5}\,\sqrt{E^2_e -
  m^2_e}\,E_e\,F(E_e, Z = 1)\,(E_0 - E_e -
\omega)^2\,\frac{1}{\omega}\int \frac{d\Omega_{\gamma}}{4\pi}
\nonumber\\
\hspace{-0.3in}&& \times \,\bigg\{\Big(1 + B_0\,
\frac{\vec{\xi}_n\cdot \vec{k}_{\bar{\nu}}}{E_{\bar{\nu}}}\Big)\Big\{
\Big[\frac{\beta^2 - (\vec{n}\cdot \vec{\beta}\,)^2}{(1 - \vec{n}\cdot
    \vec{\beta})^2} \Big(1 + \frac{\omega}{E_e}\Big) +
  \frac{\omega^2}{E^2_e}\,\frac{1}{1 - \vec{n}\cdot \vec{\beta}}\Big]
- \frac{m_e}{E_e}\Big[\frac{\beta^2 - (\vec{n} \cdot
    \vec{\beta}\,)^2}{(1 - \vec{n}\cdot \vec{\beta}\,)^2} \zeta^0_e +
  \frac{\omega}{E_e}\frac{\zeta^0_e - (\vec{n}\cdot
    \vec{\beta}\,)(\vec{n}\cdot \vec{\zeta}_e)}{(1 - \vec{n}\cdot
    \vec{\beta}\,)^2} \nonumber\\
\hspace{-0.3in}&& + \frac{\omega^2}{E^2_e} \frac{\zeta^0_e -
  \vec{n}\cdot \vec{\zeta}_e}{(1 - \vec{n}\cdot \vec{\beta}\,)^2}
\Big) \Big] \Big\} + \Big(A_0 \vec{\xi}_n + a_0
\frac{\vec{k}_{\bar{\nu}}}{E_{\bar{\nu}}}\Big) \cdot
\Big\{\Big[\frac{\beta^2 - (\vec{n} \cdot \vec{\beta}\,)^2}{(1 -
    \vec{n}\cdot \vec{\beta}\,)^2}\Big( \frac{\vec{k}_e}{E_e} +
  \vec{n}\, \frac{\omega}{E_e}\Big) + \frac{\omega}{E_e}\,
  \frac{\vec{\beta} - \vec{n}\,(\vec{n}\cdot \vec{\beta}\,)}{1 -
    \vec{n}\cdot \vec{\beta}} + \frac{\omega^2}{E^2_e}\,
  \frac{\vec{n}}{1 - \vec{n}\cdot \vec{\beta}}\Big]\nonumber\\
\hspace{-0.3in}&& - \frac{m_e}{E_e}\Big[\frac{\beta^2 - (\vec{n} \cdot
    \vec{\beta}\,)^2}{(1 - \vec{n}\cdot
    \vec{\beta}\,)^2}\,\vec{\zeta}_e +
  \frac{\omega}{E_e}\,\frac{\zeta^0_e - \vec{n}\cdot \vec{\zeta}_e}{(1
    - \vec{n}\cdot \vec{\beta}\,)^2}\,\vec{\beta} +
  \frac{\omega}{E_e}\,\frac{\vec{n}}{1 - \vec{n}\cdot
    \vec{\beta}}\,\zeta^0_e +
  \frac{\omega^2}{E^2_e}\,\frac{\vec{n}\,\big(\zeta^0_e - \vec{n}\cdot
    \vec{\zeta}_e\big)}{(1 - \vec{n}\cdot
    \vec{\beta}\,)^2}\Big\}\bigg\}.
\end{eqnarray}
The integration over the directions of the vector $\vec{n}$ we carry
out by using the results obtained in \cite{Ivanov2017}. We get
\begin{eqnarray}\label{eq:B.13}
\hspace{-0.15in}&&\frac{d^6\lambda_{\beta^-_c\gamma}(E_e,\omega,\vec{k}_e,
  \vec{k}_{\bar{\nu}}, \vec{\xi}_n, \vec{\xi}_e)}{d\omega d E_e
  d\Omega_ed\Omega_{\bar{\nu}}} = (1 + 3 g^2_A) \,
\frac{\alpha}{\pi}\,\frac{|G_V|^2}{16 \pi^5}\,\sqrt{E^2_e -
  m^2_e}\,E_e\,F(E_e, Z = 1)\,(E_0 - E_e -
\omega)^2\,\frac{1}{\omega}\bigg\{\Big(1 + B_0\,
\frac{\vec{\xi}_n\cdot \vec{k}_{\bar{\nu}}}{E_{\bar{\nu}}}\Big)
\nonumber\\
\hspace{-0.3in}&& \times \,\Big\{\Big[\Big(1 + \frac{\omega}{E_e} +
  \frac{1}{2}\,\frac{\omega^2}{E^2_e}\Big)\,\Big[\frac{1}{\beta}\,{\ell
      n}\Big(\frac{1 + \beta}{1 - \beta}\Big) - 2\Big] +
  \frac{\omega^2}{E^2_e}\Big] - \frac{m_e}{E_e}\,\zeta^0_e\Big[1 +
  \frac{1}{\beta^2}\,\frac{\omega}{E_e}\Big(1 +
  \frac{1}{2}\,\frac{\omega}{E_e}\Big)\Big]
\,\Big[\frac{1}{\beta}\,{\ell n}\Big(\frac{1 + \beta}{1 - \beta}\Big)
  - 2\Big]\Big\}\nonumber\\
\hspace{-0.3in}&& + \Big(A_0 \vec{\xi}_n + a_0
\frac{\vec{k}_{\bar{\nu}}}{E_{\bar{\nu}}}\Big) \cdot \Big\{
\frac{\vec{k}_e}{E_e}\, \Big[1 +
  \frac{1}{\beta^2}\,\frac{\omega}{E_e}\Big(1 +
  \frac{1}{2}\,\frac{\omega}{E_e}\Big)\Big]\,\Big[\frac{1}{\beta}\,{\ell
    n}\Big(\frac{1 + \beta}{1 - \beta}\Big) - 2\Big] - \vec{\zeta}_e
\frac{m_e}{E_e}\,\Big(1 -
\frac{1}{2\beta^2}\,\frac{\omega^2}{E^2_e}\Big)\,
\Big[\frac{1}{\beta}\,{\ell n}\Big(\frac{1 + \beta}{1 - \beta}\Big) -
  2\Big]  \nonumber\\
\hspace{-0.3in}&& -
\vec{\beta}\,\frac{m_e}{E_e}\,\zeta^0_e\Big[\frac{1}{
    \beta^2}\,\frac{\omega}{E_e}\Big[\frac{1}{\beta}\,{\ell
      n}\Big(\frac{1 + \beta}{1 - \beta}\Big) - 2\Big] + \frac{1}{2
    \beta^2}\, \frac{\omega^2}{E^2_e}\,\Big(\frac{3 -
    \beta^2}{\beta^2}\Big[\frac{1}{\beta}\,{\ell n}\Big(\frac{1 +
      \beta}{1 - \beta}\Big) - 2\Big] - 2\Big)\Big]\Big\}\bigg\}.
\end{eqnarray}
The r.h.s. of Eq.(\ref{eq:B.13}), rewritten in terms of irreducible
correlation structures (see Eq.(\ref{eq:1})), takes the form
\begin{eqnarray*}%\label{eq:B.14}
\hspace{-0.15in}&&\frac{d^6\lambda_{\beta^-_c\gamma}(E_e,\omega,\vec{k}_e,
  \vec{k}_{\bar{\nu}}, \vec{\xi}_n, \vec{\xi}_e)}{d\omega d E_e
  d\Omega_ed\Omega_{\bar{\nu}}} = (1 + 3 g^2_A) \,
\frac{\alpha}{\pi}\,\frac{|G_V|^2}{16 \pi^5}\,\sqrt{E^2_e -
  m^2_e}\,E_e\,F(E_e, Z = 1)\,(E_0 - E_e - \omega)^2
\nonumber\\
\hspace{-0.3in}&& \times \,\bigg\{\frac{1}{\omega}\, \Big[\Big(1 +
  \frac{\omega}{E_e} +
  \frac{1}{2}\,\frac{\omega^2}{E^2_e}\Big)\,\Big[\frac{1}{\beta}\,{\ell
      n}\Big(\frac{1 + \beta}{1 - \beta}\Big) - 2\Big] +
  \frac{\omega^2}{E^2_e}\Big] + a_0\,\frac{\vec{k}_e\cdot
  \vec{k}_{\bar{\nu}}}{E_e E_{\bar{\nu}}}\, \frac{1}{\omega}\, \Big(1
+ \frac{1}{\beta^2}\,\frac{\omega}{E_e} + \frac{1}{2
  \beta^2}\,\frac{\omega^2}{E^2_e}\Big) \nonumber\\
\hspace{-0.3in}&& \times\,\Big[\frac{1}{\beta}\, {\ell n}\Big(\frac{1
    + \beta}{1 - \beta}\Big) - 2\Big] + A_0\, \frac{\vec{\xi}_n \cdot
  \vec{k}_e}{E_e}\,\frac{1}{\omega} \, \Big(1 +
\frac{1}{\beta^2}\,\frac{\omega}{E_e}+ \frac{1}{2
  \beta^2}\,\frac{\omega^2}{E^2_e}\Big) \,\Big[\frac{1}{\beta}\, {\ell
    n}\Big(\frac{1 + \beta}{1 - \beta}\Big) - 2\Big] +
B_0\,\frac{\vec{\xi}_n \cdot
  \vec{k}_{\bar{\nu}}}{E_{\bar{\nu}}}\nonumber\\
\hspace{-0.3in}&& \times \,\frac{1}{\omega} \, \Big[\Big(1 +
  \frac{\omega}{E_e} + \frac{1}{2}\,\frac{\omega^2}{E^2_e}\Big)
  \,\Big[\frac{1}{\beta}\,{\ell n}\Big(\frac{1 + \beta}{1 -
      \beta}\Big) - 2\Big] + \frac{\omega^2}{E^2_e}\Big] + (- 1)\,
\frac{\vec{\xi}_e \cdot \vec{k}_e}{E_e}\,\frac{1}{\omega} \,\Big(1 +
\frac{1}{\beta^2}\,\frac{\omega}{E_e} + \frac{1}{2
  \beta^2}\,\frac{\omega^2}{E^2_e}\Big)\Big]\nonumber\\
\hspace{-0.3in}&& \times \,\Big[\frac{1}{\beta}\,{\ell n}\Big(\frac{1
    + \beta}{1 - \beta}\Big) - 2\Big] + (- 1)\,\frac{m_e}{E_e}\,a_0\,
\frac{\vec{\xi}_e \cdot \vec{k}_{\bar{\nu}}}{E_{\bar{\nu}}}
\,\frac{1}{\omega} \,\Big(1 -
\frac{1}{2\beta^2}\,\frac{\omega^2}{E^2_e}\Big)\,
\Big[\frac{1}{\beta}\,{\ell n}\Big(\frac{1 + \beta}{1 - \beta}\Big) -
  2\Big] + (- 1)\, \frac{m_e}{E_e}\, A_0 \nonumber\\
\hspace{-0.3in}&& \times \,\vec{\xi}_n \cdot
\vec{\xi}_e\,\frac{1}{\omega} \,\Big(1 -
\frac{1}{2\beta^2}\,\frac{\omega^2}{E^2_e}\Big)\,
\Big[\frac{1}{\beta}\,{\ell n}\Big(\frac{1 + \beta}{1 - \beta}\Big) -
  2\Big] + (- 1) \,A_0\, \frac{\vec{\xi}_n \cdot
  \vec{k}_e)(\vec{\xi}_e\cdot \vec{k}_e)}{(E_e + m_e) E_e} \,
\Big\{\frac{1}{\omega}\,\Big(1 -
\frac{1}{2\beta^2}\,\frac{\omega^2}{E^2_e}\Big)\nonumber\\ 
\hspace{-0.3in}&& \times \,\Big[\frac{1}{\beta}\,{\ell n}\Big(\frac{1
    + \beta}{1 - \beta}\Big) - 2\Big] + (1 + \sqrt{1 -
  \beta^2}\,)\Big[\frac{1}{\beta^2}\, \frac{\omega}{E_e}
  \,\Big[\frac{1}{\beta}\, {\ell n}\Big(\frac{1 + \beta}{1 -
      \beta}\Big) - 2\Big] + \frac{1}{2 \beta^2}\,
  \frac{\omega^2}{E^2_e}\, \,\Big(\frac{3 -
    \beta^2}{\beta^2}\nonumber\\
\end{eqnarray*}
\begin{eqnarray}\label{eq:B.14}   
\hspace{-0.3in}&& \times \Big[\frac{1}{\beta}\,{\ell n}\Big(\frac{1 +
    \beta}{1 - \beta}\Big) - 2\Big] - 2\Big)\Big\} +
(-1)\,a_0\,\frac{(\vec{\xi}_e \cdot \vec{k}_e)(\vec{k}_e \cdot
  \vec{k}_{\bar{\nu}})}{(E_e + m_e) E_e
  E_{\bar{\nu}}}\,\Big\{\frac{1}{\omega}\,\Big(1 -
\frac{1}{2\beta^2}\,\frac{\omega^2}{E^2_e}\Big)\Big[\frac{1}{\beta}\,{\ell
    n}\Big(\frac{1 + \beta}{1 - \beta}\Big) - 2\Big] \nonumber\\
\hspace{-0.3in}&&+ (1 + \sqrt{1 - \beta^2}\,)\Big[\frac{1}{\beta^2}\,
  \frac{\omega}{E_e} \,\Big[\frac{1}{\beta}\, {\ell n}\Big(\frac{1 +
      \beta}{1 - \beta}\Big) - 2\Big] + \frac{1}{2 \beta^2}\,
  \frac{\omega^2}{E^2_e}\, \,\Big(\frac{3 -
    \beta^2}{\beta^2}\Big[\frac{1}{\beta}\,{\ell n}\Big(\frac{1 +
      \beta}{1 - \beta}\Big) - 2\Big] - 2\Big)\Big\}\nonumber\\
\hspace{-0.3in}&& - B_0 \frac{(\vec{\xi}_n\cdot
  \vec{k}_{\bar{\nu}})(\vec{\xi}_e\cdot \vec{k}_e)}{E_e
  E_{\bar{\nu}}}\,\frac{1}{\omega} \, \Big(1 +
\frac{1}{\beta^2}\,\frac{\omega}{E_e}+ \frac{1}{2
  \beta^2}\,\frac{\omega^2}{E^2_e}\Big) \,\Big[\frac{1}{\beta}\, {\ell
    n}\Big(\frac{1 + \beta}{1 - \beta}\Big) - 2\Big] \bigg\}.
\end{eqnarray}
It is seen that the terms with the correlation structures
$(\vec{\xi}_n\cdot \vec{\xi}_e)(\vec{k}_e \cdot \vec{k}_{\bar{\nu}})$
and $(\vec{\xi}_n\cdot \vec{k}_e)(\vec{\xi}_e \cdot
\vec{k}_{\bar{\nu}})$, inducing the correlation coefficients $S(E_e)$
and $U(E_e)$, respectively, do not appear in the electron-energy and
angular distribution of the neutron radiative beta decay for polarized
neutrons, polarized electrons, unpolarized protons and unpolarized
photons. This confirms the results, obtained in Appendix A, that there
are no contributions of the radiative corrections of order
$O(\alpha/\pi)$, caused by one-virtual photon exchanges, to the
correlation coefficients $S(E_e)$ and $U(E_e)$, respectively.


\newpage

\begin{thebibliography}{9}
\bibitem{Jackson1957a} J. D. Jackson, S. B. Treiman, and H. W. Wyld
  Jr., {\it Possible tests of time reversal invariance in beta decay},
  Phys. Rev. {\bf 106}, 517 (1957); \\ DOI:
  https://doi.org/10.1103/PhysRev.106.517.

\bibitem{Ebel1957} M. E. Ebel and G. Feldman, {\it Further remarks on
  Coulomb corrections in allowed beta transitions}, Nucl. Phys. {\bf
  4}, 213 (1957); \\ DOI:
  https://doi.org/10.1016/0029-5582(87)90020-4.

\bibitem{Ivanov2021} A. N. Ivanov, R. H\"ollwieser, N. I. Troitskaya,
  M. Wellenzohn, and Ya. A. Berdnikov, {\it On the correlation
    coefficient $T(E_e)$ of the neutron beta decay, caused by the
    correlation structure invariant under discrete P, C and T
    symmetries}; arXiv: 2101.01014 [hep-ph].

 \bibitem{Feynman1958} R. P. Feynman and M. Gell--Mann, {\it Theory of
  Fermi interaction}, Phys. Rev.{\bf 109}, 193 (1958); \\ DOI:
   https://doi.org/10.1103/PhysRev.109.193.

 \bibitem{Marshak1969} R. E. Marshak, Riazuddin, and C. P. Ryan, in
   {\it Theory of weak interactions in particle physics},
   Wiley-Interscience, A Division of John Wiley $\&$ Sons, Inc. New
   York, p. 41 (1969).
  
\bibitem{Abele2008} H. Abele, {\it The neutron. Its properties and
  basic interactions}, Progr. Part. Nucl. Phys. {\bf 60}, 1 (2008);
  \\ DOI: https://doi.org/10.1016/j.ppnp.2007.05.002.

\bibitem{Nico2009} J. S. Nico, {\it Neutron beta decay}, J. Phys. G:
  Nucl. Part. Phys. {\bf 36}, 104001 (2009); \\ DOI:
  https://doi.org/10.1088/0954-3899/36/10/104001.

\bibitem{Abele2018} B. M\"arkisch, H. Mest, H. Saul, X. Wang,
  H. Abele, D. Dubbers, M. Klopf, A. Petoukhov, C. Roick, T. Soldner,
  and D. Werder, {\it Measurement of the weak axial-vector coupling
    constant in the decay of free neutrons using a pulsed cold neutron
    beam}, Phys. Rev. Lett. {\bf 122}, 242501 (2019); \\ DOI:
  https://doi.org/10.1103/PhysRevLett.122.242501; arXiv: 1812.04666
  [nucl-ex].

\bibitem{Sirlin2018} A. Czarnecki, W. J. Marciano, and A. Sirlin, {\it
  Neutron lifetime and axial coupling constant}, Phys. Rev. Lett. {\bf
  120}, 202002 (2018); \\ DOI:
  https://doi.org/10.1103/PhysRevLett.120.202002; arXiv: 1802.01804
  [hep-ph].

\bibitem{PDG2020} P. A. Zyla {\it et al.}, {\it Review of particle
  physics} (Particle Data Group), Prog. Theor. Exp. Phys. {\bf 2020},
  083C01 (2020); \\ DOI: https://doi.org/10.1093/ptep/ptaa104.



\bibitem{DeAlfaro1973} V. De Alfaro, S. Fubini, G. Furlan, and
  C. Rossetti, in {\it Currents in hadronic physics}, Noth--Holland
  Publishing Company Amsterdam $\cdot$ London, American Elsevier
  Publishing Company, Inc.  New York 1973.
  
\bibitem{Sirlin1967} A. Sirlin, {\it General properties of the
  electromagnetic corrections to the beta decay of a physical
  nucleon}, Phys. Rev. {\bf 164}, 1767
  (1967); \\ DOI:https://doi.org/10.1103/PhysRev.164.1767.

\bibitem{Shann1971} R. T. Shann, {\it Electromagnetic effects in the
  decay of polarized neutrons}, Nuovo Cimento A {\bf 5}, 591
  (1971).\\ DOI: https://doi.org/10.1007/BF02734566.

\bibitem{Ando2004} S. Ando, H. W. Fearing, V. Gudkov, K. Kubodera,
  F. Myhrer, S. Nakamura, and T. Sato, {\it Neutron beta--decay in
    effective field theory}, Phys. Lett. B {\bf 595}, 250 (2004); DOI:
  https://doi.org/10.1016/j.physletb.2004.06.037.
  
\bibitem{Gudkov2006} V. Gudkov, G. I. Greene, and J. R. Calarco, {\it
  General classification and analysis of neutron beta-decay
  experiments}, Phys. Rev. C {\bf 73}, 035501 (2006); \\ DOI:
  https://doi.org/10.1103/PhysRevC.73.035501.

\bibitem{Ivanov2013} A. N. Ivanov, M. Pitschmann, and
  N. I. Troitskaya, {\it Neutron $\beta$--decay as a laboratory for
    testing the standard model}, Phys. Rev. D {\bf 88}, 073002 (2013);
  \\ DOI: https://doi.org/10.1103/PhysRevD.88.073002; arXiv:1212.0332
     [hep--ph].

\bibitem{Wilkinson1970} D. H. Wilkinson and B. E. F. Macfield, {\it
  The numerical evaluation of radiative corrections of order $\alpha$
  to allowed nuclear $\beta$--decay}, Nucl. Phys. A {\bf 158}, 110
  (1970); DOI: https://doi.org/10.1016/0375-9474(70)90055-2. 

 \bibitem{Ivanov2017} A. N. Ivanov, R. H\"ollwieser, N. I. Troitskaya,
   M. Wellenzohn, and Ya. A. Berdnikov, {\it Precision analysis of
     electron energy spectrum and angular distribution of neutron beta
     decay with polarized neutron and electron}, Phys. Rev. C {\bf
     95}, 055502 (2017); \\ DOI: 10.1103/PhysRevC.95.055502;
   arXiv:1705.07330 [hep-ph].


\bibitem{Ivanov2019a} A. N. Ivanov, R. H\"ollwieser, N. I. Troitskaya,
  M. Wellenzohn, and Ya. A. Berdnikov, {\it Test of the Standard Model
    in neutron beta decay with polarized electrons and unpolarized
    neutrons and protons}, Phys. Rev. D 99, 053004 (2019); \\ DOI:
  10.1103/PhysRevD.99.053004; arXiv:1811.04853 [hep-ph].

\bibitem{Bilenky1959} S. M. Bilen'kii, R. M. Ryndin,
  Ya. A. Smorodinskii, and Ho Tso-Hsiu, {\it On the theory of the neutron
  beta decay}, JETP {\bf 37}, 1759 (1959) (in Russian);
  Sov. Phys. JETP, {\bf 37}(10), 1241 (1960).

 \bibitem{Wilkinson1982} D. H. Wilkinson, {\it Analysis of neutron beta
  decay}, Nucl. Phys. A {\bf 377}, 474 (1982); \\ DOI:
   https://doi.org/10.1016/0375-9474(82)90051-3.

\bibitem{Itzykson1980} C. Itzykson and J.--B. Zuber, in {\it Quantum
  field theory}, McGraw--Hill Inc., New York, 1980.
  

\bibitem{Jackson1957b} J. D. Jackson, S. B. Treiman, and H. W. Wyld
  Jr., {\it Coulomb corrections in allowed beta transitions},
  Nucl. Phys. {\bf 4}, 206 (1957); DOI:
  https://doi.org/10.1016/0029-5582(87)90019-8.


\bibitem{Jackson1958} J. D. Jackson, S. B. Treiman, and H. W. Wyld,
  Jr., {\it Note on relativistic coulomb wave functions},
  Z. Phys. {\bf 150}, 640 (1958); DOI:
  https://doi.org/10.1007/BF01340460. 

\bibitem{Callan1967} C. G. Callan and S. B. Treiman, {\it Electromagnetic
  simulation of T violation in beta decay}, Phys. Rev. {\bf 162}, 1494
  (1967); \\ DOI: https://doi.org/10.1103/PhysRev.162.1494.
  
\bibitem{Ando2009} S. I. Ando, J. A. McGovern, and T. Sato, {\it The D
  coefficient in neutron beta decay in effective field theory},
  Phys. Lett. B {\bf 677}, 109 (2009); \\ DOI:
  https://doi.org/10.1016/j.physletb.2009.04.088.

  
\bibitem{Fierz1937} M. Fierz, {\it Zur Fermischen Theorie des
  $\beta$-Zerfalls}, Z. Physik {\bf 104}, 553 (1937); \\ DOI:
  https://doi.org/10.1007/BF01330070.

  
\bibitem{Hardy2020} J. C. Hardy and I. S. Towner, {\it Superallowed
  $0^+ \to 0^+$ nuclear beta decays: 2020 critical survey, with
  implications for $V_{ud}$ and CKM unitarity}, Phys. Rev. C {\bf
  102}, 045501 (2020); \\ DOI:
  https://doi.org/10.1103/PhysRevC.102.045501.

\bibitem{Severijns2019} M. Gonz\'alez--Alonso, O. Naviliat--Cuncic,
  and N. Severijns, {\it New physics searches in nuclear and neutron
    beta decay}, Prog. Part. Nucl. Phys. {\bf 104}, 165 (2019);
  \\ DOI: https://doi.org/10.1016/j.ppnp.2018.08.002.

\bibitem{Abele2019} H. Saul, Ch. Roick, H. Abele, H. Mest, M. Klopf,
  A. Petukhov, T. Soldner, X. Wang, D. Werder, and B.  M\"arkisch,
  {\it Limit on the Fierz interference term b from a measurement of
    the beta asymmetry in neutron decay}, Phys. Rev. Lett. {\bf 125},
  112501 (2020); \\ DOI:
  https://doi.org/10.1103/PhysRevLett.125.112501.

\bibitem{Young2019} V. Cirigliano, A. Garcia, D. Gazit,
  O. Naviliat-Cuncic, G. Savard, and A. Young, {\it Precision beta
    decay as a probe of new physics}, arXiv:1907.02164 [nucl-ex].

 \bibitem{Ivanov2019y} A. N. Ivanov, R. H\"ollwieser, N. I. Troitskaya,
  M. Wellenzohn, and Ya. A. Berdnikov, {\it Neutron dark matter decays
    and correlation coefficients of neutron beta decays},
  Nucl. Phys. B {\bf 938}, 114 (2019); \\ DOI:
  https://doi.org/10.1016/j.nuclphysb.2018.11.005.
  

\bibitem{Sun2020} X. Sun {\it et al.}, {\it Improved limits on Fierz
  interference using asymmetry measurements from the ultracold neutron
  asymmetry (UCNA) experiment} (UCNA Collaboration), Phys.  Rev. C
  {\bf 101}, 035503 (2020); \\ DOI:
  https://doi.org/10.1103/PhysRevC.101.035503.

\bibitem{Ivanov2020} A. N. Ivanov, R. H\"ollwieser, N. I. Troitskaya,
  M. Wellenzohn, and Ya. A. Berdnikov, {\it Precision analysis of
    pseudoscalar interactions in neutron beta decays}, Nucl. Phys. B
  {\bf 951}, 114891 (2020); \\ DOI:
  https://doi.org/10.1016/j.nuclphysb.2019.114891; arXiv:1905.04147
  [hep-ph].

\bibitem{Weinberg1958} S. Weinberg, {\it Charge symmetry of weak
  interactions}, Phys. Rev. {\bf 112}, 1375 (1958); \\ DOI:
  https://doi.org/10.1103/PhysRev.112.1375.

\bibitem{Bodek2016} K. Bodek, {\it Beta-decay correlations in the LHC
  era}, Acta Phys. Polon. B {\bf 47}, 349 (2016); \\ DOI:
  10.5506/APhysPolB.47.349.


\bibitem{Berman1958} S. M. Berman, {\it Radiative corrections to muon
  and neutron decay}, Phys. Rev. {\bf 112}, 267 (1958); \\ DOI:
  https://doi.org/10.1103/PhysRev.112.267.

\bibitem{Kinoshita1959}
 T. Kinoshita and A. Sirlin, {\it Radiative
  corrections to Fermi interactions}, Phys. Rev. {\bf 113}, 1652
  (1959); \\ DOI: https://doi.org/10.1103/PhysRev.113.1652
    
\bibitem{Berman1962} S. M. Berman and A. Sirlin,{\it Some
  considerations on the radiative corrections to muon and neutron
  decay}, Ann. Phys.  (N.Y.)  {\bf 20}, 20 (1962); \\ DOI:
  https://doi.org/10.1016/0003-4916(62)90114-8.


\bibitem{Kaellen1967} G. K\"all${\acute{\rm e}}$n, {\it Radiative
  corrections to beta decay and nucleon form factors}, Nucl. Phys. B
  {\bf 1}, 225 (1967); \\ DOI:
https://doi.org/10.1016/0550-3213(67)90125-3.


\bibitem{Abers1968} E. S. Abers, D. A. Dicus, R. E. Norton, and
  H. R. Queen, {\it Radiative corrections to the Fermi part of
    strangeness-conserving beta decay}, Phys. Rev. {\bf 167}, 1461
  (1968); \\ DOI: https://doi.org/10.1103/PhysRev.167.1461.

\bibitem{Herczeg2001} P. Herczeg, {\it Beta decay beyond the standard
  model}, Progr. Part. Nucl. Phys. {\bf 46}, 413 (2001); \\ DOI:
  https://doi.org/10.1016/S0146-6410(01)00149-1.

  
\bibitem{Severijns2006} N. Severijns, M. Beck, and O. Naviliat-Cuncic,
  {\it Tests of the standard electroweak model in nuclear beta decay},
  Rev. Mod. Phys. {\bf 78}, 991 (2006); \\ DOI:
  https://doi.org/10.1103/RevModPhys.78.991.

\bibitem{Lee1956a} T. D. Lee and C. N. Yang, {\it Charge conjugation,
  a new quantum number $G$ , and selection rules concerning a nucleon
  anti-nucleon system}, Nuovo Cimento {\bf 10}, 749 (1956); \\ DOI:
  https://doi.org/10.1007/BF02744530.
  

\bibitem{Gardner2001} S. Gardner and C. Zhang, {\it Sharpening
  low-energy, Standard-Model tests via correlation coefficients in
  neutron beta decay}, Phys. Rev. Lett. {\bf 86}, 5666 (2001); \\ DOI:
  https://doi.org/10.1103/PhysRevLett.86.5666.

\bibitem{Gardner2013} S. Gardner and B. Plaster, {\it Framework for
  maximum likelihood analysis of neutron beta decay observables to
  resolve the limits of the V - A law}, Phys. Rev. C {\bf 87}, 065504
  (2013); The contribution to 4th International Conference on Particle
  Physics and Astrophysics (ICPPA-2018) 22–26 October 2018, Moscow,
  Russian Federation;\\ DOI: https://doi.org/10.1103/PhysRevC.87.065504.

\bibitem{Ivanov2018} A. N. Ivanov, R. H\"ollwieser, N. I. Troitskaya,
  M. Wellenzohn, and Ya. A. Berdnikov, {\it Tests of the standard
    model in neutron beta decay with polarized neutron and electron
    and an unpolarized proton}, Phys. Rev. D 98, 035503 (2018);
  \\ DOI: 10.1103/PhysRevD.99.053004; arXiv:1805.03880 [hep-ph].

\bibitem{Bhattacharya2012} T. Bhattacharya, V. Cirigliano,
  S. D. Cohen, A. Filipuzzi, M. Gonz\'alez-Alonso, M. L. Graesser,
  R. Gupta, and Huey-Wen Lin, {\it Probing novel scalar and tensor
    interactions from (ultra)cold neutrons to the LHC}, Phys. Rev. D
  {\bf 85}, 054512 (2012); \\ DOI:
  https://doi.org/10.1103/PhysRevD.85.054512.

\bibitem{Gardner2012} S. Gardner, V. Cirigliano, P. Fierlinger,
  C. Fischer, K. Jansen, S. Paul, F. J. Llanes-Estrada, H. W. Lin, and
  W. M. Snow, {\it Round Table: Resolving Physics BSM at Low
    Energies}, PoS ConfinementX (2012) 024; \\ DOI:
  https://doi.org/10.22323/1.171.0024.

\bibitem{Cirigliano2012} V. Cirigliano, M. Gonz\'alez-Alonso,
M. L. Graesser, {\it Non-standard charged current interactions: beta
  decays versus the LHC}, JHEP {\bf 2012}, 25 (2012); \\ DOI:
https://doi.org/10.1007/JHEP10(2012)025.

\bibitem{Cirigliano2013a} V. Cirigliano and M. J. Ramsey-Musolf, {\it
  Low energy probes of physics beyond the Standard Model},
  Prog. Part. Nucl. Phys. {\bf 71}, 2 (2013); \\ DOI:
  https://doi.org/10.1016/j.ppnp.2013.03.002.

\bibitem{Cirigliano2013b} V. Cirigliano, S. Gardner, and B. Holstein,
  {\it Beta decays and non-standard interactions in the LHC era},
  Prog. Part. Nucl. Phys. {\bf 71}, 93 (2013); \\ DOI:
  https://doi.org/10.1016/j.ppnp.2013.03.005.
  
\bibitem{Hardy2009}J. C. Hardy and I. S. Towner, {\it Superallowed
  $0^+ \to 0^+$ nuclear beta decays: A new survey with precision tests
  of the conserved vector current hypothesis and the standard model},
  Phys. Rev. C {\bf 79}, 055502 (2009); \\ DOI:
  https://doi.org/10.1103/PhysRevC.79.055502.
  
\bibitem{Blatt1952} J. M. Blatt and V. F. Weisskopf, {\it Theoretical
  nuclear physics}, John Wily $\&$ Sons, New York 1952.

\bibitem{Antognini2013} A.  Antognini {\it et al.},
  {\it Proton structure from the measurement of 2S-2P transition
    frequencies of muonic hydrogen}, Science {\bf 339} 417 (2013);
  \\ DOI: 10.1126/science.1230016.
  
 \bibitem{Ivanov2020b} A. N. Ivanov, R. H\"ollwieser, N. I. Troitskaya,
  M. Wellenzohn, and Ya. A. Berdnikov, {\it Corrections of order
    $O(E^2_e/m^2_N)$, caused by weak magnetism and proton recoil, to
    the neutron lifetime and correlation coefficients of the neutron
    beta decay}, Results in Physics {\bf 21}, 103806 (2021); \\ DOI:
  https://doi.org/10.1016/j.rinp.2020.103806; arXiv: 2010.14336
  [hep-ph].


\bibitem{Ivanov2017b} A. N. Ivanov, R. H\"ollwieser, N. I. Troitskaya,
  M. Wellenzohn, and Ya. A. Berdnikov, {\it Precision theoretical
    analysis of neutron radiative beta decay to order
    $O(\alpha^2/\pi^2)$}, Phys. Rev. D {\bf 95}, 113006 (2017);
  \\ DOI: https://doi.org/10.1103/PhysRevD.95.113006.

\end{thebibliography}
\end{document}